# The Causal Effect of Transport Infrastructure: Evidence from a New Historical Database[*]

This version: August 28, 2021

Erik Lindgren,[♥] Per Pettersson-Lidbom,[♠] and Björn Tyrefors[♦]

**Abstract**

In this paper, we analyze the effect of transport infrastructure investments in railways. As a testing ground, we use data from a new historical database that includes *annual* panel data on approximately 2,400 Swedish *rural* geographical areas during the period 1860-1917. We use a staggered event study design that is robust to treatment effect heterogeneity. Importantly, we find extremely large reduced-form effects of having access to railways. For real nonagricultural income, the cumulative treatment effect is approximately 120% after 30 years. Equally important, we also show that our reduced-form effect is likely to reflect growth rather than a reorganization of existing economic activity since we find no spillover effects between treated and untreated regions. Specifically, our results are consistent with the big push hypothesis, which argues that simultaneous/coordinated investment, such as large infrastructure investment in railways, can generate economic growth if there are strong aggregate demand externalities (e.g., Murphy et al. 1989). We used plant-level data to further corroborate this mechanism. Indeed, we find that investments in local railways dramatically, and independent of initial conditions, increase local industrial production and employment on the order of 100-300% across almost all industrial sectors.

[*] We would like to thank Clément de Chaisemartin, Xavier D'Haultfoeuille, Dave Donaldson and Rick Hornbeck for their useful comments. This work was financed by an ERC consolidator grant (616496) (Pettersson-Lidbom).

[♥] Department of Economics, Stockholm University; email: erik.lindgren@ne.su.se
[♠] Department of Economics, Stockholm University and Research Institute of Industrial Economics (IFN), email: pp@ne.su.se
[♦] Research Institute of Industrial Economics (IFN) and Department of Economics, Stockholm University and IFN, email: bjorn.tyrefors@ifn.se.

# 1. Introduction

Transportation infrastructure is often considered a key to promoting growth and development (e.g., Banerjee et al. (2020)). Indeed, Rostow (1960, p. 302) argues that "the introduction of the railroad has been historically the most powerful single initiator of take-offs". However, it has been difficult to substantiate this claim. Indeed, Fogel (1964) found little evidence for the historical importance of railways to economic growth. Additionally, recent quasi-experimental work on the direct effects of railroads on income and industrialization has reached a similar conclusion since most of the estimated direct effects of transport infrastructure are small, i.e., approximately 6% (see Section 6.2.1 in Redding and Turner (2015)).[1]

Nonetheless, this conclusion is perhaps premature since it is very challenging to estimate the causal effects of railways on economic activity, as discussed by Donaldson (2015), among others. Indeed, Donaldson argued that a successful empirical analysis must overcome problems of noncomparability between treatment and control regions and ensure that the stable unit treatment value assumption (SUTVA) assumption holds, i.e., that there are no spillover effects across regions. There is also an additional econometric problem that must be addressed, namely, that the standard two-way fixed effect (TWFE) estimator (i.e., linear regression with period and group fixed effects) can be severely biased when the treatment effect is heterogeneous (e.g., Borusyak et al. (2021), Callaway and Sant'Anna (2020), de Chaisemartin and d'Haultfoeuille (2020, 2021), Godman and Bacon (2021), Sun and Abraham (2020)). Indeed, this is likely a serious problem for studies of the effect of transport infrastructure on economic activity since many of them have used TWFE estimators, as discussed by Redding and Turner (2015) and Redding (2020), and that "transport improvements are likely to have heterogeneous treatment effects" (Redding (2020, p. 27).[2]

In this paper, we can address all of these empirical problems. Specifically, we employ the new event study estimator developed by de Chaisemartin and d'Haultfoeuille (2020, 2021) to circumvent the problem with the TWFE estimator when treatment effects are heterogeneous.

An equally important contribution is that we can implement an unusually credible and statistically powerful event study design since we collected and digitized highly spatially disaggregated data on an *annual* basis for the universe of approximately 2,400 *rural* geographical areas (i.e., local governments) in Sweden for the period of 1860-1917. Indeed, our event study design was based on a maximum of 1,340 openings of new local railways. Moreover, having access to panel data with a very large number of *yearly* observations also makes it possible to convincingly rule out any violation of the parallel trend assumption. Indeed, we find that the treatment and control groups have similar trends, at least 25 years before the railways started to be built. Another attractive feature of having data measured on an annual basis in a staggered treatment design is that all the dynamic treatment effects will be appropriately specified. On the other hand, if the data are recorded on, say, a decennial basis, the dynamic effects will be improperly measured since some units may have been treated for one year, whereas others may have been treated for 9 years.

---

[1] There are a number of recent studies that estimates the direct effect of railways on economic activity, for example, Atack and Margo (2011), Banerjee et al. (2020), Berger and Enflo (2017), Berger (2019), Andersson et al (2020), and Donaldson (2018).
[2] Importantly, in our Appendix, we show that the results from some prominent recent historical studies of the effect of railways on economic activity, i.e., Donaldson (2018), Donaldson and Hornbeck (2016), Heblich et al. (2020), and Hornung (2015) are not robust to treatment effects heterogeneity.

Another key contribution is that we can measure the potential effect of transport infrastructure on the local economy with much more precision and in much greater detail than in any previous study since our historical database includes highly spatially disaggregated data on arguably the most important measures of local economic activity, i.e., real income in the nonagricultural sector, agricultural land values, and population size.

In fact, with our historical database, it is even possible to make a credible test of one of the prosed mechanisms behind the effect of investments in railways on economic growth, i.e., the "big push" hypothesis, as discussed by Rosenstein-Rodan (1943) and Murphy, Shleifer and Vishny (1989), among others. The basic argument for a big push development strategy is based on the fact that the size of the domestic market is too small for firms to generate enough sales to make adoption of increasing returns technologies profitable. However, a simultaneous/coordinated investment (a big push), such as large infrastructure investment in railways, can generate economic growth if there are strong aggregate demand externalities. To test the big push hypothesis, we digitized plant-level data for the universe of manufacturing firms that was part of the Swedish manufacturing census for the years 1913-1917. With our plant-level data, we are able to measure even highly localized agglomeration effects.[3]

We find extremely large effects of transport infrastructure on local economic activity Indeed, real income in the nonagricultural sector is increased by approximately 120% over a 30-year period after investment in local railways. Thus, this effect is approximately 20 times larger than that in many previous studies (see Section 6.2.1 in Redding and Turner (2015)). We also find strong support for the big push hypothesis since investments in local railways increase both local industrial output and employment in almost all industrial sectors in the range of 100-300%.

---
[3] See, for example, Duranton and Kerr (2015) on the importance of having access to plant-level data when analyzing agglomeration economies and Rosenthal and Strange (2020) for using the appropriate level of aggregation.

Moreover, we also show that initial economic conditions, i.e., the historical level of economic activity, do not seem to matter much for the size of the effect of investments in railways on either real income in the nonagricultural sector or industrial production and employment. We interpret this as additional support for the big push theory since these findings are consistent with expectations, the willingness of firms to invest depends on their expectations of other firms to invest, determines the choice of equilibrium, rather than history (i.e., initial conditions), as discussed by Krugman (1991).

We also find that railways have an impact on agricultural land values and population size. However, these effects are much smaller than the effect on local business activity., i.e., both of which increase by approximately 15% over a 30-year period after the opening of a railway. The finding of much smaller effects on land values and population size also seems consistent with the big push theory since it reasonable to argue that local investments in railways had the largest effect on factors directly related to the local industrial sector.

Another key contribution of our study is that we empirically distinguish growth from reorganization using an idea suggested by Redding and Turner (2015, Section 4.3), i.e., "with panel data, one could estimate the change in the treated region following a change in transportation costs and the change in the untreated region following the change in the treated region". Importantly, the suggested test by Redding and Turner is also a test for violation of SUTVA since it tests for spillover effects across treatment and control regions.[4] Reassuringly, we find little evidence of spillover effects when estimating the placebo effects of our staggered treatment design where adjacent neighboring regions without access to railways act as placebo treatment groups following a change in railway access in the treated regions. Indeed, the estimated placebo effects are rather precisely estimated zeros. Finding little evidence of either positive or negative spillover effects also suggests that our reduced-form effects reflect growth rather than a reorganization of existing economic activity, as discussed by Redding and Turner (2015). Additionally, the lack of any spillover effects seems consistent with the big push theory of expectation-driven demand externalities. Thus, if there are no investments in railways, then it seems reasonable to expect that firms do not expect other firms to invest.

---

[4] If SUTVA-violating spillovers between regions are negative, then the true treatment effect will be overstated. Similarly, if the spillovers are positive, then standard estimators will understate the true effect (Donaldson 2015).

This paper adds to the literature on the effect of transport infrastructure investments in general and on railways in particular.[5] Specifically, this study is the first to use a credible event study design that solves the problem with heterogeneous treatment effects.[6] Moreover, the highly spatially disaggregated historical data from Sweden are also considerably better and more comprehensive than the data in any previous study since we have annual data on both a very large number of railway openings and the key measures of local economic activity.

This paper also adds to the literature on economies of agglomeration and large push infrastructure investments (e.g., Rosenstein-Rodan (1943), Murphy et al. (1989), Kline and Moretti (2014)). In fact, we are the first to perform an empirical test for aggregate demand externalities due to the large push mechanism.[7]

Relatedly, we add to the literature on the relative importance of expectations and history in determining equilibrium (e.g., Krugman (1991) and Matsuyama (1991)) by constructing a novel empirical test. Indeed, our empirical test is distinct from other tests of spatial persistence and path dependence in economics (e.g., Allen and Donaldson (2020), Davis and Weinstein (2002), and Bleakely and Lin (2012)).[8]

---

[5] See Redding and Turner (2015) and Redding (2020) on infrastructure investments in general. On investments in railways, see Donaldson (2018), Donaldson and Hornbeck (2016), Heblich et al. (2020), and Hornung (2015). There are also three recent Swedish studies of the effect of railways. Berger and Enflo (2017) investigated whether urban development, i.e., population size, differs depending on having access to a state-owned railway using data from 81 small urban areas, i.e., towns. The second study by Berger (2019) analyzed the long-run local effect of having access to a state-owned railway line on population size and employment types using data from only two years: 1850 and 1900. The third study by Andersson et al. (2020) analyzed the effect of railways on innovative activity. Importantly, these studies did not analyze all of the available railway openings. Indeed, Enflo and Berger only analyzed the openings of 22 state-owned railways, while Berger used 168 openings of state-owned railways. In sharp contrast, we use all available openings, including all private railways, since it is otherwise impossible to solve the problem with heterogenous treatment effects. Moreover, the Swedish studies did not use annual data or analyze the effect of railways on real income or land values.

[6] To the best of our knowledge, Heblich et al. (2020) and Hornung (2015) are the only studies using an event study design. However, Heblich et al. (2020) used a conventional event study design, in which they imposed unnecessary restrictions on the dynamics, i.e., binned endpoints, and their approach is unreliable if the treatment effect is heterogeneous across units or time. Moreover, their event study is not based on annul data but only on decennial data. Similarly, Hornung (2015) used a conventional event study design with unnecessary restrictions on the dynamic event study specification. His event study was also based on triennial data. All other studies of the effects of railways on economic outcomes used a "static" TWFE approach, i.e., Anderson et al. (2020), Atack and Margo (2011), Berger (2019), Berger and Enflo (2017, Donaldson (2018), and Donaldson and Hornbeck (2016). Importantly, we find that the results from many of these studies are not robust to heterogeneity in the treatment effect (see Appendix).

[7] Rodríguez-Clare (1997, p. 103) writes regarding the existence of aggregate demand spillovers "Unfortunately, there are not many empirical studies of this phenomenon." He also notes that no study providence evidence for the big push mechanism. Similarly, Lin and Rauch (2020) also notes the lack of evidence from self-fulfilling expectations.

[8] Allen and Donaldson (2020) provide a theoretical framework for interpreting the results from these empirical tests of spatial persistence and path dependence. However, their model does not consider forward looking behavior, i.e., the mechanism we analyze.

This paper is the first to be able to construct a convincing empirical test that can distinguish growth from reorganization based on the observed effects of infrastructure to the best of our knowledge.[9] Moreover, this empirical test is also a novel approach for detecting violations of SUTVA, which has received little attention in the literature on market integration (Donaldson 2015).

Finally, our result that real nonagricultural income was so greatly affected by having access to railways speaks not only to the literature on the effect of transport infrastructure investments on economic activity but also to the literature on the effect of political institutions on economic development more broadly.[10] Specifically, in our previous work, Lindgren et al. (2019, 2021), we argued that changes in local political institutions played a key role in the Swedish growth miracle, i.e., Sweden's transformation from one of the poorest countries in Europe in the mid-19$^{th}$ century to one of the richest countries worldwide in the 1960s. Indeed, we argue that the introduction of the new weighted voting system at the local level in 1862, which gave industrialists considerable political power to influence the decisions made at municipal town meetings, such as decisions on building local railways, was a fundamental determinant of economic growth. In this paper, we provide evidence that investments in local railways were indeed one of the engines of growth.

The rest of this paper is structured as follows. Section 2 describes the construction and financing of the Swedish railway network. Section 3 presents the data. Section 4 discusses the empirical design: an event study with staggered treatment adoption. Section 5 presents the results, while Section 6 describes our test to distinguish growth from reorganization. Section 7 tests the mechanism, i.e., the big push theory. Finally, Section 8 discusses and concludes.

---

[9] On this point, Redding and Turner (2015) note that "The existing reduced form literature generally does not provide a basis for separately identifying the two effects".
[10] See, for example, Bogart (2020) for a recent survey of the role of institutions in shaping infrastructure outcomes.

## 2. The Swedish railway network

Sweden started to build railways in 1856, which for Europe was very late (e.g., Hedin 1967). Remarkably, however, Sweden had one of the largest railway networks worldwide in 1914. For example, there were 25 kilometers of railroads for every 10,000 inhabitants, a figure more than twice that of any European country. Only four countries (the US, Canada, Australia, and Argentina) had larger networks in per capita terms (Hedin 1967).

Another interesting feature of the Swedish railway network is that 75% of railways were built, financed and operated by private interests, while the remaining 25% were built by the state (Ottosson and Anderssson Skog 2013). Figure 1 shows how the length of state and local railways developed from 1856 to 1917. This dual feature of the Swedish railway network stems from a decision made by the Swedish Parliament in 1854 that the railway network should be constructed and financed only partly by the state.[11] Indeed, the state should only be responsible for the main (trunk) lines, while all local branches should be built, financed and operated by private local interests. As a result, Sweden had a dual system of railway network ownership, planning and control, whereas many other countries had only one of these two systems: private (e.g., the US and Great Britain) or state (e.g., France and Prussia) railway networks. As a result, the Sweden railway network was to a large extent both unstructured and uncoordinated due to the large influence of local private interests (Ottosson and Anderssson Skog (2013, p. 16), Nicander (1980, p. 8)). Indeed, after the first decision in 1854, there was an ongoing debate in Parliament over the relative merits of having privately owned or state-owned railway networks, as discussed by Oredsson (1969). Thus, this conflict of interest resulted in Sweden having a curious mixture of state and private railway lines until 1939, when the total railway network was nationalized.[12]

---

[11] On the planning of state-owned trunk lines, see the discussion in Anderssson Skog (1993).
[12] For a discussion of the decision to nationalize the privately-owned railways, see Anderssson Skog (1993).

Figure 1. Length of privately owned and state-owned railways

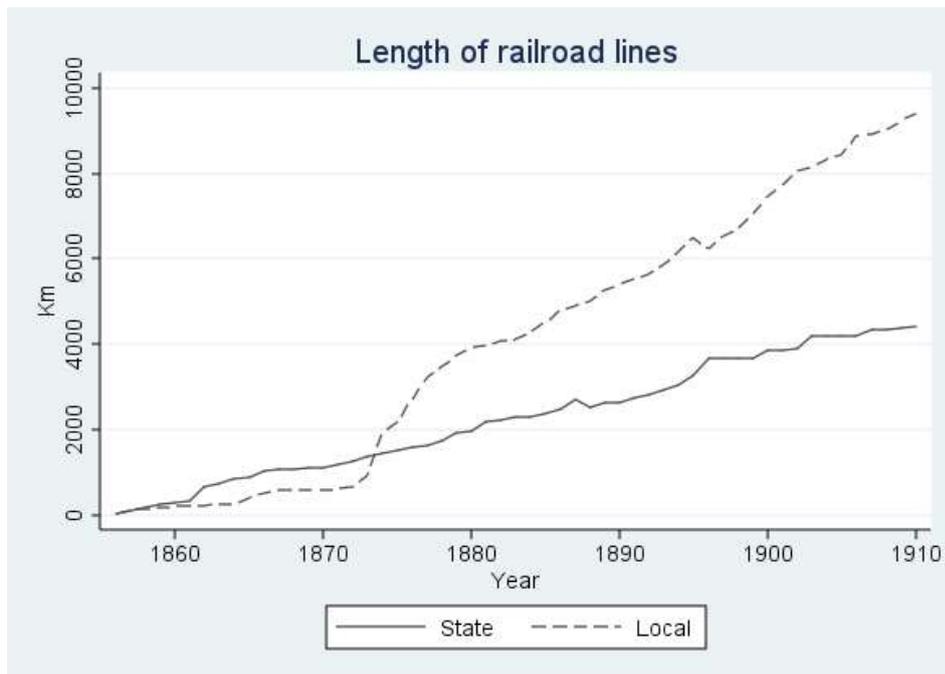

With regard to the building of private local railways, the following procedure was required.[13] First, a group of initiators, i.e., the local private interests, conducted an investigation led by an experienced railway technician. The technician then drafted a plan for the construction of the local railway together with an estimate of its costs. Second, the local private interests applied for a railway license from the central government, where the investigation by the technician provided the necessary documentation. The application was then reviewed by the Road and Water Construction Board ("Väg och Vattenbyggnadsstyrelsen"), which was a government agency overseeing investments in transport infrastructure. Third, if the application was granted, then the group of initiators formed a limited liability company that would be in charge of the construction and operation of the local railway. Importantly, as discussed by Ottoson and Andersson Skog (2013, p. 15), the state had little influence on these private companies once the application had been granted. In fact, the state always approved the application if 2/3 of the required capital had been procured (e.g., Andersson-Skog, (1993, p. 38.) Fourth, before the construction of the local railway could start, the limited liability company was required to have acquired all the necessary funding.

The financing of the state-owned trunk lines was predominately carried out via foreign loans that were procured by the National Debt Office. At the beginning of the 20th century, the national debt due to these investments in state railways constituted nearly 90% of GDP. However, the financing of private local railways was almost exclusively raised within the country using several different sources, i.e., issuing stocks (40%) and taking out bond loans (30%), central government loans (20%), and promissory note loans and other short-term loans (10%). Notably, state aid constituted only 20% of all private local railway capital.

---
[13] This description is based on Nicander (1980).

Additionally, local governments were often of decisive importance for the construction of local railways since they were usually large stockholders. Indeed, they were often the largest stockholder (e.g., Oredsson (1989) and Hedin (1969)). Swedish local governments could invest in local railways because of the Local Government Act of 1862 since it explicitly gave local governments permission to deal with all economic matters of local importance, including local infrastructure investments; previously, local governments had been in charge of primary education, poor relief and matters related to the clergy. All local government spending was to be financed via a proportional income tax rate that local governments could set completely freely.[14] At the beginning of the 20th century, the average income tax rate was nearly 10% but could be as high as 40%. Local government investments in local railways were predominately financed via long-term loans, typically with a 40-year maturity, which required approval from the central government. During the period 1870-1908, more than 700 local governments applied for such loans.

The local financing of railways was highly controversial during this period. The reason was that the decisions to finance local railways were made at regular municipal town meetings where only a few voters could have the majority of votes due to the weighted voting system where voters received votes in proportion to their taxable income, as discussed in our previous research, i.e., Lindgren et al. (2019, 2020). Mellquist (1974, p 139-155) analyzes in close detail how railway investment decisions were made in 128 distinct cases. He finds that in almost all of his investigated cases, local private interests, typically in the form of a few industrialists, could determine the outcome of the decisions to financially support local railways at these municipal town meetings since they usually possessed many more votes than all other meeting attendants combined. Moreover, Mellquist finds that most landowning farmers voted against the financing of local railways. Thus, this finding clearly illustrates the existence of a historical social conflict between landowners and industrialists regarding the attitude toward industrialization and the institutional organization of the labor market in the Swedish setting.

---

[14] The state grants constituted less than 10% of local government revenues.

For example, having access to local railways would make it more difficult for landowners to control their labor force. Another and perhaps the most important reason for landowners' opposition to railways is that they were afraid of losing their political power.[15] Indeed, if industrialists had more to gain from local railways than landowners, then the Swedish weighted voting system based on taxable income implied that landowners would also receive fewer votes in the future than industrialists. We also obtain this finding, as discussed further below.

Turning to a description of the distribution of railway openings in local governments over time, i.e., the source of variation on which our event study design is based, we present Figure 2, which shows the years of opening of state railways across all local governments during the period 1856-1917. There were a total of 267 such openings, and most of them occurred in approximately 1860 when the main lines were built. However, during the same period, there were a total of 1,073 private railway openings, and there were relatively few openings before 1870. Thus, out of the 1,340 openings of new Swedish railways up to 1917, 80% of them were private railways.

As a result of this preponderance of private investments in railways, local railways transported many more people and goods than state-owned railways. For example, in 1910, 38 million people traveled on local lines, but only 20 million traveled on trunk lines. The tonnage of goods transported by local railways was also 2 times higher than that transported by state-owned railways.

---

[15] For a general description of the historical social conflict between landowners and industrialists, see Acemoglu et al. (2005).

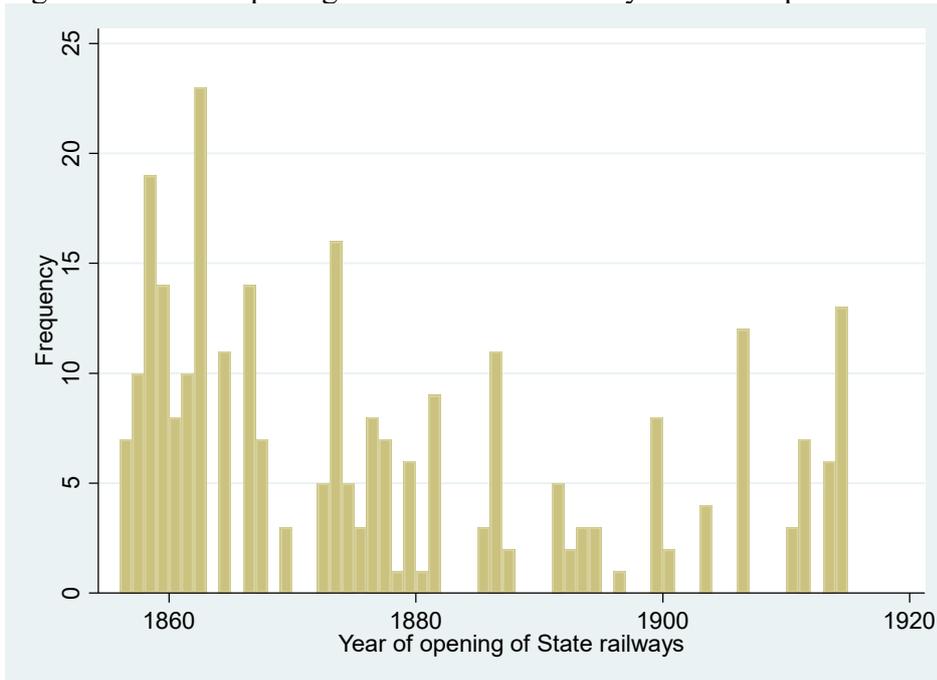

Figure 2. Year of opening of state-owned railways in municipalities

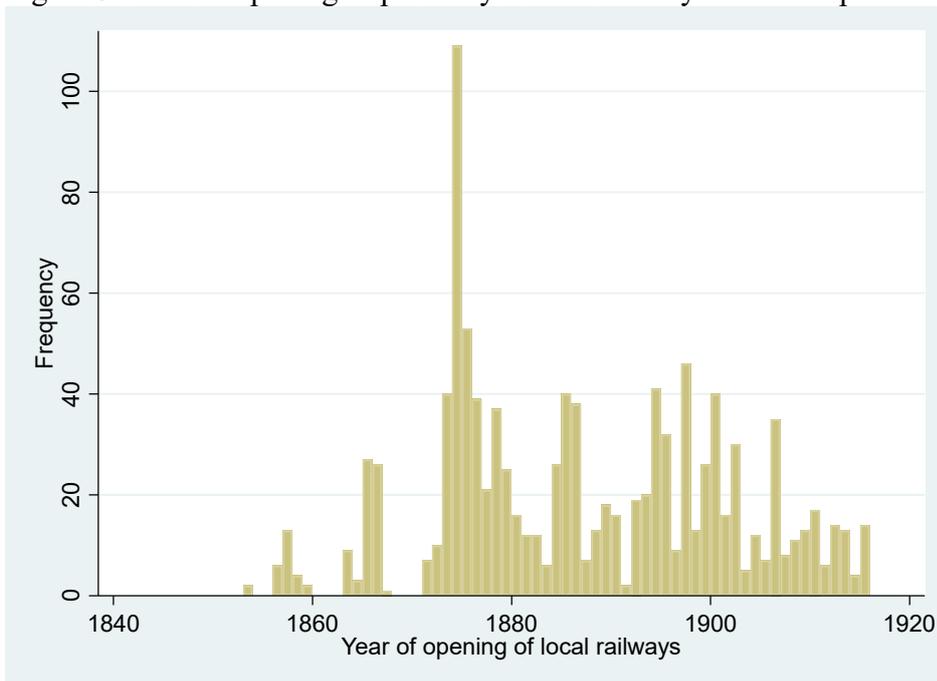

Figure 3. Year of opening of privately-owned railways in municipalities

# 3. Data

We collected highly disaggregated data on an annual basis for approximately 2,400 rural local governments for the period 1860-1917.[16] We have also collected plant-level data for the universe of all firms included in the Swedish manufacturing census for the years 1913-1917. We will discuss the local government data in this section, while the plant-level data will be discussed when we test the big push theory in Section 7.

Before, we discuss the local government data, we briefly describe the unit of analysis, i.e., Swedish rural local governments, and the rationale for focusing on them in our analysis. Local governments are the lowest government level in Sweden. There were approximately 2,500 local governments during the period of investigation, and almost all of them were situated in rural areas. However, 101 were classified as cities in 1916. These urban jurisdictions are excluded from the analysis due to reasons of comparability since Swedish towns/cities are very heterogeneous along many dimensions. In sharp contrast, rural local governments are highly comparable since they are governed by exactly the same set of state rules. Indeed, they have exactly the same set of tax instruments and fiscal responsibilities. In addition to the issue of comparability, we also focus on rural areas since we are interested in testing the theory of the big push, which concerns how underdeveloped areas (i.e., typically rural areas) can escape development traps by a big push strategy. Thus, it seems natural to exclude urban areas in this case.

It is still important to stress that most people in Sweden lived in rural areas during the period of investigation, e.g., 90% in 1850 and 72% in 1917. Furthermore, most of the industrialization also occurred in rural areas. For example, as late as 1913, 58% of the total employment in the industrial sector was based in rural areas and 54% of the total production value. The average geographical size of a rural government was 182 square kilometers (≈13×13 km), with a standard deviation of 81,230. The average population of a rural local government was 1,714 in 1917, with a standard deviation of 2,012. Thus, these rural local governments were sparsely populated areas with an average population density of 9 inhabitants per square kilometer. Nonetheless, the rural local governments were still of substantial economic importance since they were in charge of primary schooling and poor relief and made very large investments in local railways, as discussed earlier. Again, it is important to stress that 90% of the tax revenues were raised at the local level via a proportional tax rate that they could set with complete discretion.

---

[16] The data collection was supported by Per Pettersson-Lidbom's ERC consolidator grant.

Turning to the local government data, we will analyze three outcomes: population size, real nonagricultural income and agricultural land values.

We digitized yearly data on the population sizes of the municipalities at the end of the year. These data were originally collected by Statistics Sweden and come from two sources: for the period 1860-1874, they come from unpublished material from the Swedish National Archives ("summariska folkmängdsredogörelserna"). For the period 1874-1917, the population data are from the publication "Kommunernas fattigvård och finanser 1874-1917" (BiSOS U).

We also digitized data on nonagricultural income for the period 1880-1917 from the publication "Kommunernas fattigvård och finanser 1874-1917" (BiSOS U). Here, nonagricultural income included both labor income and all other types of capital income, including corporate income. Most importantly, the rules governing the Swedish income tax system were identical across all local governments since they were based on the exact same rules: the Central Government Tax Act implemented in 1862 (SFS 1861:34 Bewillningsförordning 1862). As a result, with our data, we can avoid problems in comparing real income both across jurisdictions and across time.

Additionally, we digitized data on the assessed value of agricultural land for the period 1880-1917 from the publication "Kommunernas fattigvård och finanser 1874-1917" (BiSOS U). The Central Government Tax Act also governed real estate taxation. Specifically, the taxation of agricultural land was based on the assessed property value, which generally reflected the true value market value, as discussed by Lindgren (2017). Indeed, an investigation by Flodström (1912) found that the assessed agricultural property value was approximately 90% of the sales value. Importantly, these assessments occurred on a regular basis, usually every third year. Table 1 presents the descriptive statistics of these three outcomes.

Table 1. Summary statistics of the outcome data

| Dependent variable | Mean | St. Dev | Number of observations |
|---|---|---|---|
| Real nonagricultural income | 8,905 | 32,114 | 86,703 |
| Agricultural land values | 6,404 | 5,609 | 86,645 |
| Population size | 1,651 | 1,843 | 137,806 |

Note: The data for real nonagricultural income and land values cover the period 1881-1916, while the population data cover the period 1860-1917.

## 4. Empirical design

Our empirical research design is a DID design with staggered treatment adoption. Importantly, we define our treatment indicator based on when the railway was started to be built rather than opened to ensure that the no-anticipation assumption is fulfilled.[17] No anticipation requires that a group's current outcome do not depend on its future treatments and is a key assumption for an event study design (e.g., de Chaisemartin and d'Haultfoeuille (2021), Sun and Abraham (2020)).

We can now define the population regression model as follows:

(1) $\quad Y_{it} = \alpha_i + \lambda_t + \beta_{it} Rail_{it} + v_{it}$,

where $i$ denotes a region, i.e., the geographical area of a local government, and $t$ is time. $\alpha_i$ is a region fixed effect, and $\lambda_t$ is a year fixed effect. $Y_{it}$ is a measure of real nonagricultural income, agricultural land values, or population size. All three outcomes will be expressed in logarithmic form since this convention is used in the literature on transport infrastructure (e.g., equation 33 in Redding (2020)). $Rail_{it}$ is an indicator variable that takes the value of 1 if region $i$ has started to build a railway in year $t$.

Most importantly, the population model allows for heterogeneous treatment effects by region and time and, thus, by the number of years since treatment. The model also incorporates the parallel trends assumption, whereby the expected outcome absent from the treatment is $\alpha_i + \lambda_t$. Importantly, the parallel trends assumption is much weaker than assuming that the treatment, i.e., railways, is randomly assigned. In fact, railways are rarely built randomly; thus, the parallel trends assumption seems much more sensible in regard to investments in transport infrastructure.

---

[17] How to define the date at which the treatment occurs is an important specification issue when estimating the effect of transportation infrastructure on economic activity that has received little discussion. One possibility is to define treatment based on the year when the railway is opened for use. However, since infrastructure projects usually take a long time to complete, local economic activity can be affected even long before the opening of a railway leading to a violation of the no anticipation assumption. Indeed, as we argue in Section 7, economic activity is influenced via aggregate demand externalities, i.e., a firm's future expectations of other firms' investments decisions. Thus, it seems reasonable to arguably that local economic activity is affected already when the railway was started to be built. To investigate this issue, we collected data on the building time of all local railways, finding that 94% of such railways had a building time of at most 2 years. Thus, it seems reasonable to expect that the treatment effect might occur 2 years prior to opening. Indeed, this is also what we find.

However, estimating equation (1) as a traditional event study, i.e., through ordinary least squares (OLS) with two-way fixed effects and some lags and leads of treatment, produces estimates that are biased in the presence of heterogeneity in the treatment effects, as discussed previously. Thus, we will use an event study estimator that is robust to heterogeneous or dynamic treatment effects. There are a number of recent estimators that are robust to treatment effect heterogeneity. We have chosen to use de Chaisemartin and d'Haultfoeuille (2020, 2021), which produces identical estimates of dynamic treatment effects as Callaway and Sant'Anna (2020) if the control group is the same and without covariates.[18]

The estimand identifies the treatment effect on the switchers at the time they switch. In staggered designs, the instantaneous treatment effect is equal to the average of simple DID estimands and compares groups treated for the first time at $t$ and not yet treated at $t$, from $t-1$ to $t$. Similarly, the various estimands of the dynamic treatment effects of switchers are also defined as the weighted average of simple DID estimands. For example, the average cumulative treatment effect of switchers after one period, i.e., in period $t+1$, compares the evolution of the outcome from $t-1$ to $t+1$ between groups that are treated at period t and groups that are still untreated at period $t+1$.

The estimands of the placebo treatment effects used for testing the plausibility of the parallel trends assumption are also defined correspondingly, as discussed by de Chaisemartin and d'Haultfoeuille (2021). Specifically, de Chaisemartin and d'Haultfoeuille (2021) argue that "the long-difference placebos may lead to a more powerful test of common trends" since "the long-difference placebos test if common trends holds over several periods".

In our analysis, we estimate these treatment and placebo effects using the did_multiplegt Stata command developed by de Chaisemartin and d'Haultfoeuille (2020, 2021).

Turning to the number of switchers that identify the various dynamic and placebo treatment effects, we present Figures 4 and 5, which show the number of switchers, i.e., the number of municipalities with local railway constructions, that identify the treatment and placebo effects in our data. Importantly, the data put restrictions on the maximum number of placebo and dynamic effects it is possible to estimate. For example, when we use data for the period 1880-1916, it is only possible to estimate dynamic treatment effects for a maximum of 35 years. Moreover, in such a case the number of switchers that identifies the long-run dynamic treatment effect then also becomes increasingly smaller. For this reason, we have chosen to report dynamic effects up to 30 years after the first switch and estimate placebo effects up to 25 years before the first switch, as to avoid problems with too few switchers as further discussed below. Most, importantly, this choice has no impact on our reported results since all the dynamic treatment effects are estimated independently from each other as discussed by de Chaisemartin and d'Haultfoeuille.

---

[18] The test of parallel trends differs between de Chaisemartin and d'Haultfoeuille and Callaway and Sant'Anna. Specifically, de Chaisemartin and d'Haultfoeuille use a long-differences approach.

We start with the population data, which cover the period 1860-1917. Figure 4 reveals that the instantaneous treatment effect is based on 1,188 switchers, while the dynamic effect after 30 years is based on 654 switchers. Similarly, the first placebo effect is based on 1,183 switchers, while the long-difference placebos after 25 years are based on 586 switchers.

For the two other outcomes, real income and agricultural land values, we have fewer switchers since these data cover only the shorter time period 1880-1916. Figure 5 shows the number of switchers depending on the event time. The instantaneous treatment effect is estimated using 678 switchers, while the dynamic effect after 30 years is based on 145 switchers. The first placebo effect is based on 658 switchers, while the long-difference placebo after 25 years is based on 164 switchers.

Figure 4. Number of switchers for population size

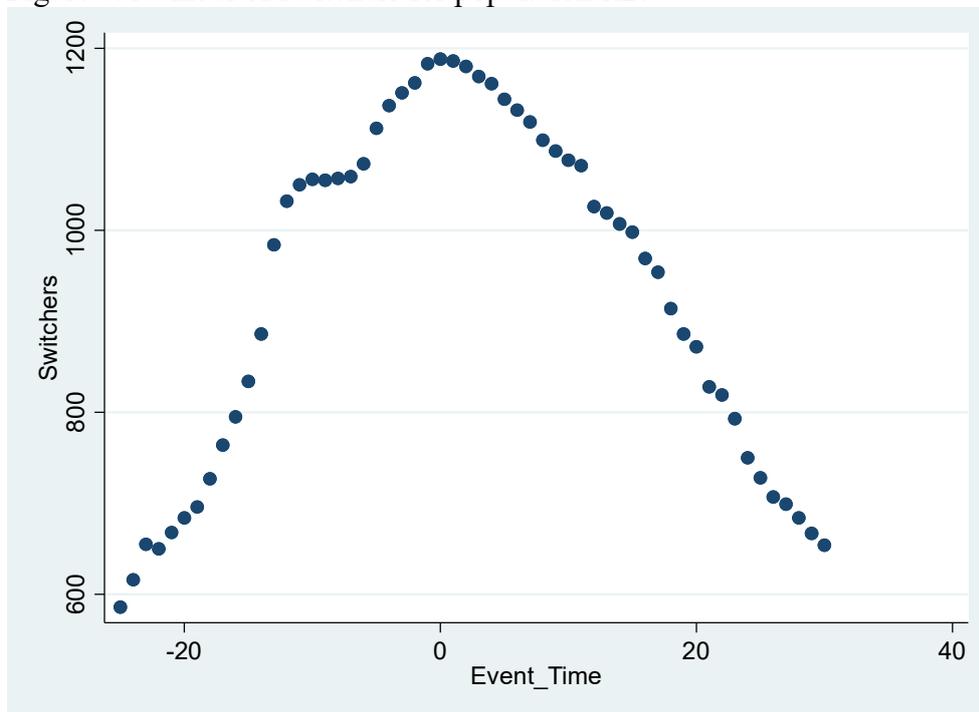

Figure 5. Number of switchers for real income and agricultural land values

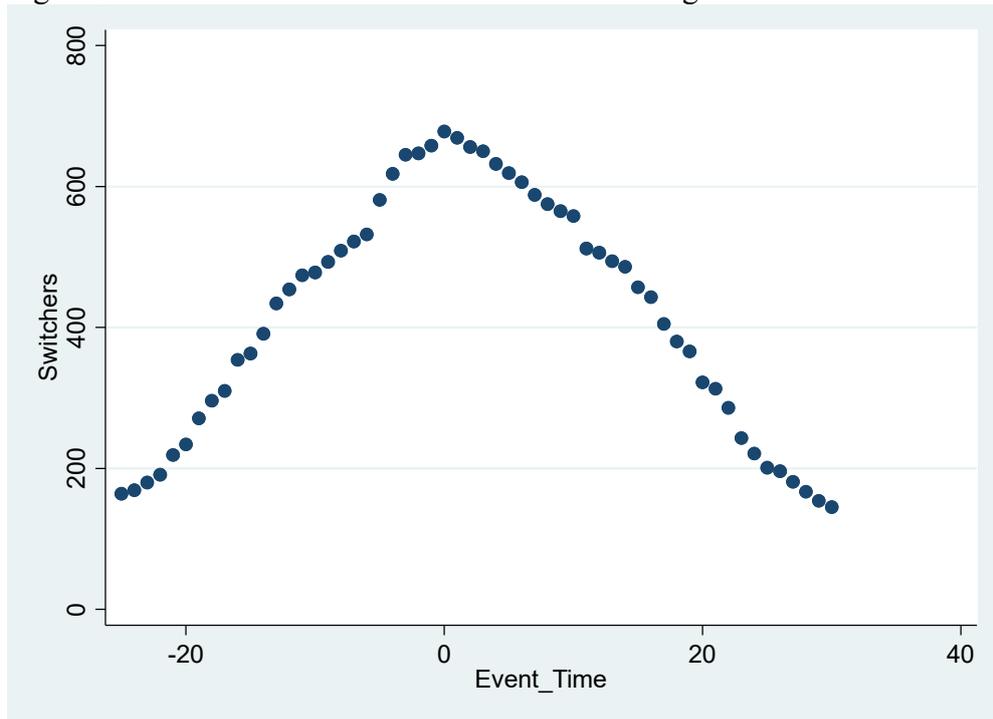

# 5. Results

In this section, we present the results from our DID design with staggered treatment adoption. We use the estimators developed by de Chaisemartin and d'Haultfoeuille (2020, 2021), which are robust to heterogeneous or dynamic treatment effects.

Starting with the effect of railways on the logarithm of real nonagricultural income, Figure 6 shows event study graphs of the relationship between the opening of a railway and the evolution of real income. Importantly, the treatment timing is based on the year in which the railways started to be built, i.e., two years prior to opening, to avoid problems with the violation of anticipation of treatment, as discussed above.[19]

Most importantly, Figure 6 shows that the treatment and control groups had remarkably similar trends 25 years before the treatment occurred. Equally important, Figure 6 shows that the dynamic treatment effects start to continuously grow directly after the treatment occurs. Thus, after 30 years, the cumulative treatment effect increased as much as 123% (=exp(0.80)-1) relative to the control group. In other words, this finding indicates that the average yearly growth in real income was 2.7% for the treatment group.

Turning to the effect of railways on logarithmic agricultural land values, Figure 7 displays the event study plot for land values. Again, the parallel trends assumption seems to hold for 25 years prior to the opening of a railway since all the placebo effects are close to zero and are not significantly different from zero.[20] The cumulative treatment effect is also much smaller for land values than real nonagricultural income since it has increased by only 16% (=exp(0.148)-1) 30 years after the opening of a railway relative to the control group. Thus, the effect of railways on land values is considerably smaller, by a factor of 7, than the effect on real income in the nonagrarian sectors of the local economy.[21]

Turning to the effect of railways on logarithmic population size, Figure 8 displays the event study plot for population size. These graphs show that the treatment and control groups have remarkably similar trends 25 years before the treatment occurs.[22] Nonetheless, the cumulative treatment effect is of the same magnitude as that for land value since the population has increased by only 17% (=exp(0.153)-1) 30 years after treatment relative to the control group.

---

[19] If the treatment is to defined when the railways is opened instead, then the treatment effect already occurs two years prior to the treatment suggesting that the no-anticipation assumption is violated during the time when a railway is being constructed. However, again this is to be expected since our proposed mechanism—aggregate demand externalities— is likely kick in already when the railway is started to being built rather when it is finished as noted above.

[20] Interestingly, we find no anticipation effect of agricultural land values, which is most likely because the assessment of land values occurs only every third year, and this variable cannot therefore detect any changes in expectation occurring at smaller time intervals than 3 years.

[21] Interestingly, this result underscores that land values miss substantial economic gains, as discussed by Hornbeck and Rotemberg (2019).

Figure 6. Event study design: the effect of railways on (log) real nonagricultural income

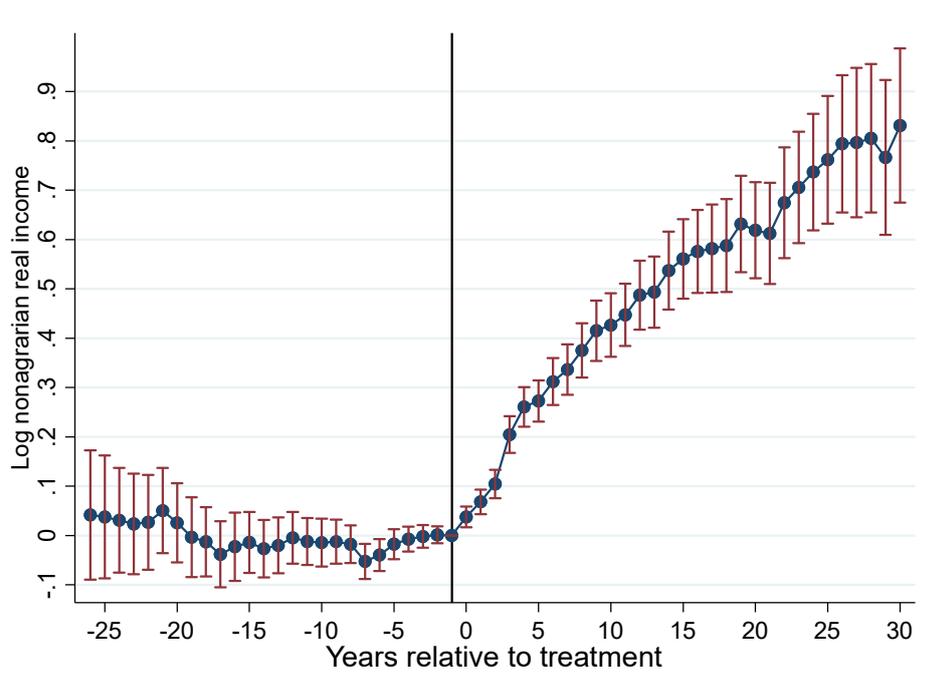

Note: The event study design is implemented using the estimator developed by de Chaisemartin and d'Haultfoeuille (2020, 2021) and the did_multiplegt Stata command.

---

[22] Interestingly, some of the estimates of the leads are still significantly different from zero despite that there are very close to zero. However, our panel data set is extremely big and therefore trivial differences in pretrends will lead to rejection of parallel trends.

Figure 7. Event study design: the effect of railways on (log) agricultural land values

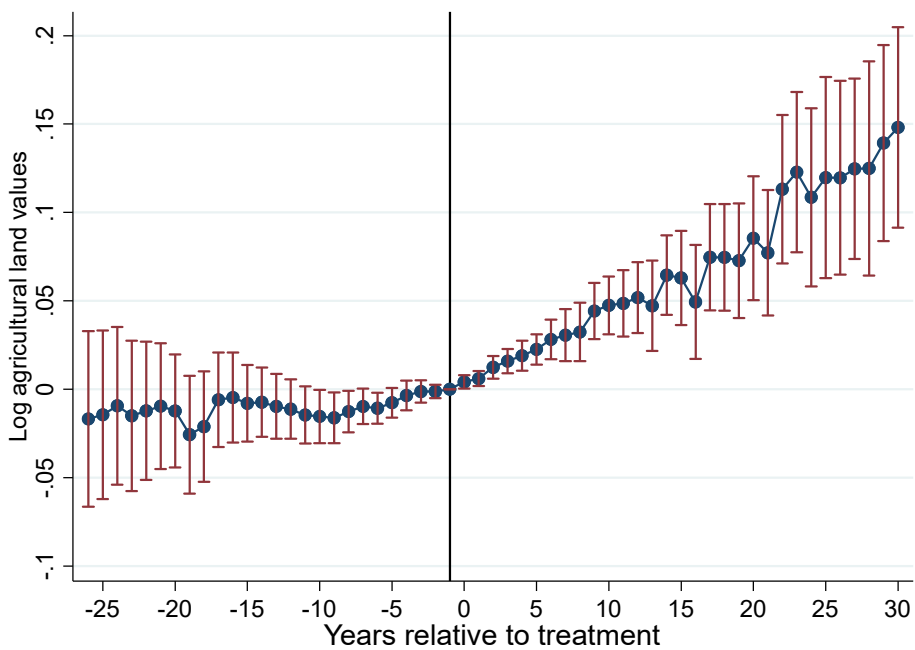

Note: The event study design is implemented using the estimator developed by de Chaisemartin and d'Haultfoeuille (2020, 2021) and the did_multiplegt Stata command.

Figure 8. Event study design: the effect of railways on (log) population size

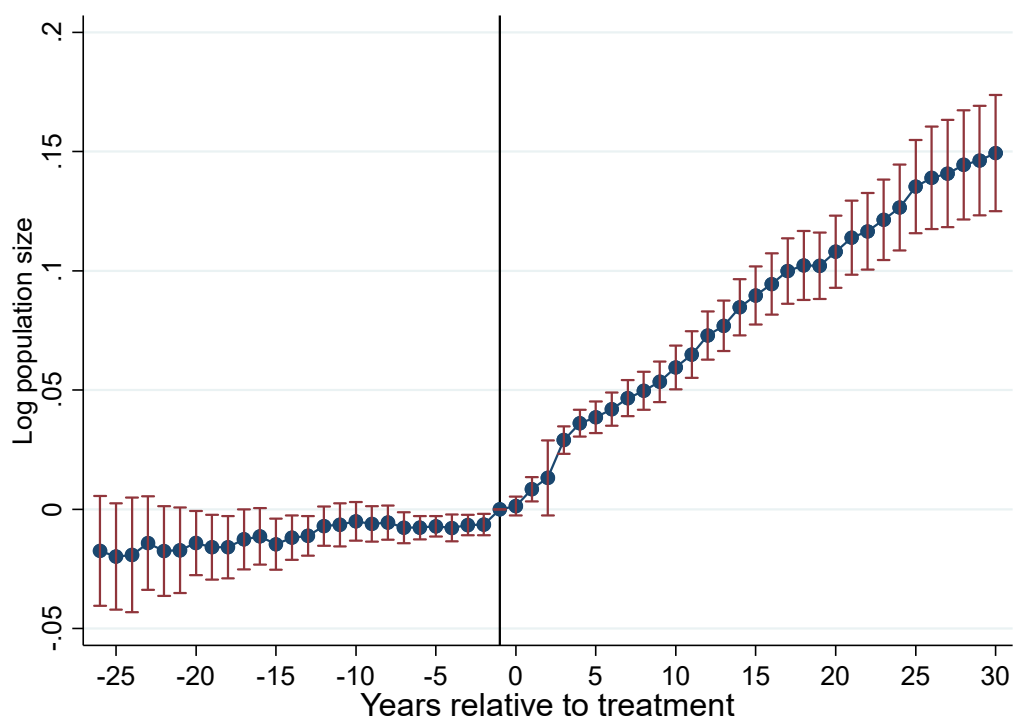

Note: The event study design is implemented using the estimator developed by de Chaisemartin and d'Haultfoeuille (2020, 2021) and the did_multiplegt Stata command.

# 6. Distinguishing growth from reorganization

In this section, we analyze whether railways changed the amount of economic activity or reorganized existing economic activity. As discussed by Redding and Turner (2015), "Determining the extent to which the observed effects of infrastructure reflect changes in the level of economic activity versus a reorganization of existing activity is fundamental to understanding the effects of infrastructure and to policy analysis".

To empirically separate growth from reorganization, we use a strategy proposed by Redding and Turner (2015) since they suggest that "with panel data, one could estimate the change in the treated region following a change in transportation costs and the change in the untreated region following the change in the treated region".

Indeed, we estimate whether there are any positive or negative spillover effects from treated to untreated regions since we use panel data. To implement our empirical test, we first restricted the sample to never-treated units, i.e., those local governments that never had access to a railway during the sample period. We then matched these never-treated units to their closest geographical neighbors with access to a railway sometime during the sample period. It is now possible to implement exactly the same type of DID specification as that in equation (1) but with the redefined treatment indicator, i.e.,

(2)  $Y_{it} = \alpha_i + \lambda_t + \pi_{it} Rail\_neigbors_{it} + v_{it}$,

where $i$ denotes a geographical area that *never* obtained access to a railway during the sample period. *Rail_neigbors$_{it}$* is now an indicator function that takes the value of 1 if a neighboring area, i.e., sharing the same border, has access to a railway at time $t$.

As a result, the $\pi_{it}$ parameter captures the spillover effects of the treatment on areas without railways. Therefore, a statistical test of $\pi_{it}=0$ is a test of whether the railway changed the amount of economic activity or reorganized existing economic activity. It also tests whether there is a SUTVA violation.[23] Specifically, finding that $\pi_{it} \neq 0$ casts doubt on our staggered DID design in equation (1) since the control group is also affected by the treatment.

We start by showing the number of switchers on which the estimation of the placebo event study is based since we must exclude all treated regions that cannot be matched with an untreated region, i.e., a neighboring region without a railway. Figure 9 displays the number of switchers when using real nonagricultural income and agricultural land values as the dependent variables. There was a dramatic reduction in the number of switchers in the placebo test compared to the previous event study design displayed in Figure 5. For example, the maximum number of switchers drops from 678 to 279, which is a reduction of 60%. Thus, we can implement the placebo event study design only for a shorter event window: only 15 years before and 20 years after the event. There is also a similar reduction in the number of switchers when population size is the dependent variable.

---

[23] The market access approach was developed by Donaldson and Hornbeck (2016) to take these spillover effects into account.

Figures 10-12 present the event study results from equation (2). To facilitate comparison with our previous event study results shown in Figures 6-10, we used the same scaling on the y-axes. We also used the year of railway opening to define the date at which the treatment occurred. All three figures reveal that there are no spillover effects of the treatment since all the placebo estimates and all the dynamic effects are close to zero and are not significantly different from zero. Thus, these results suggest that our reduced-form effects in Section 5 reflect growth rather than a reorganization of existing economic activity, as discussed by Redding and Turner (2015).

As a result, the lack of spillover effects from local railways in the Swedish setting suggests that local markets are not well integrated. One potential reason for this finding is that the local governments were such sufficiently large spatial units that transport and trade mostly took place between the villages within municipalities. Indeed, the average distances that people and goods were transported on these local railways were only approximately 18 km and 38 km, respectively, suggesting that the effect on the local economy was very localized. A second potential reason is that Sweden had a very labor-repressive agricultural system that severely restricted the movements of labor in rural areas, as discussed by Lindgren et al. (2019). A third reason for the lack of spillover effects, which we argue is the most plausible one, is that coordinated investments in local railways gave rise to aggregate demand externalities. We discuss and test this mechanism in the next section.

Figure 9. Number of "placebo" switchers

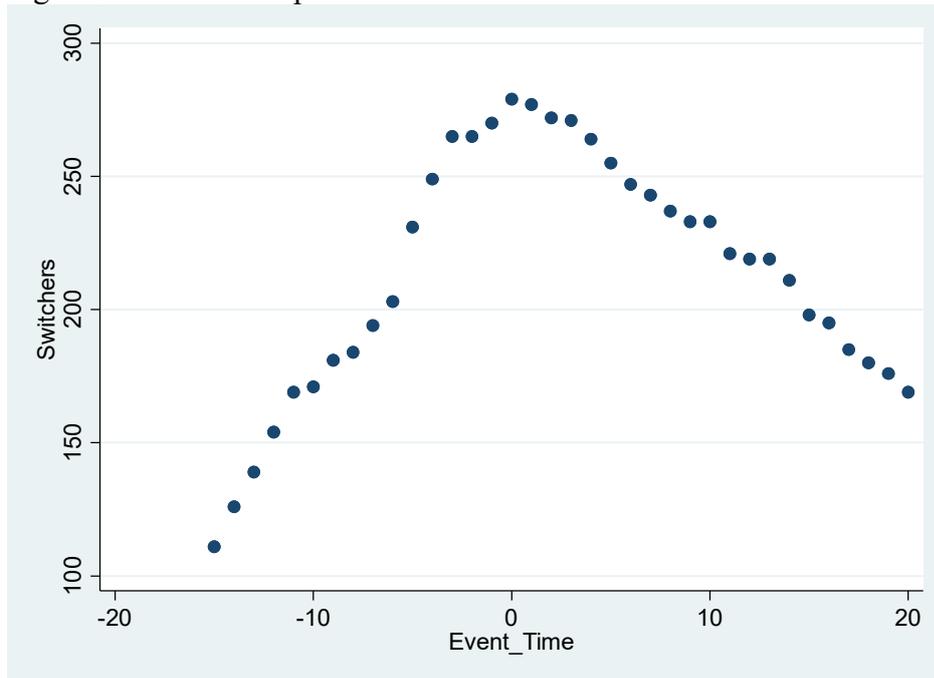

Figure 10. Event study design: the placebo effect of railways on (log) real nonagricultural income

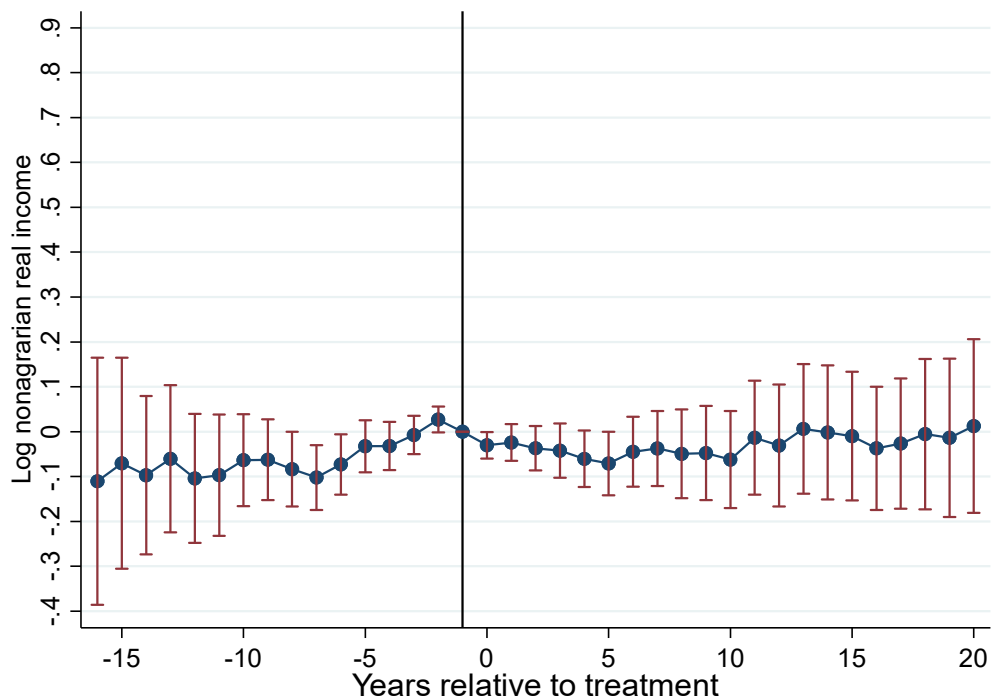

Note: The event study design is implemented using the estimator developed by de Chaisemartin and d'Haultfoeuille (2020, 2021) and the did_multiplegt Stata command.

Figure 11. Event study design: the placebo effect of railways on (log) agricultural land values

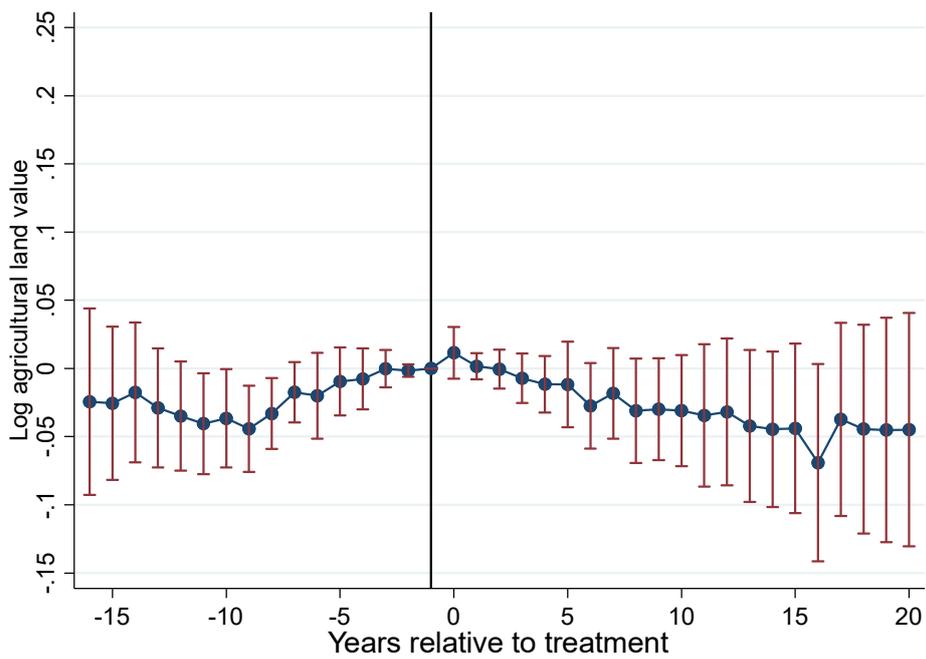

Note: The event study design is implemented using the estimator developed by de Chaisemartin and d'Haultfoeuille (2020, 2021) and the did_multiplegt Stata command.

Figure 12. Event study design: the placebo effect of railways on (log) population size

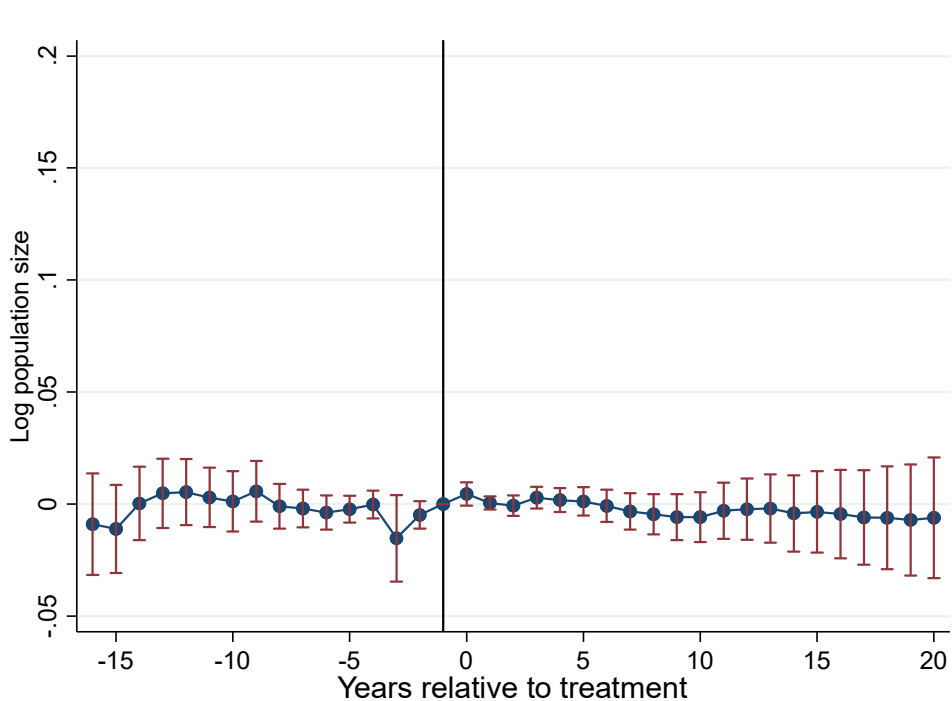

Note: The event study design is implemented using the estimator developed by de Chaisemartin and d'Haultfoeuille (2020, 2021) and the did_multiplegt Stata command.

# 7. The mechanism

In this section, we analyze one of the mechanisms that might be at work in our setting in creating the very large effects of investments in local railways on local economic activity, i.e., 120% after 30 years.

One such mechanism is the idea of a "big-push" strategy as first put forward by Rosenstein-Rodan (1943) and later formalized by Murphy et al. (1989). The idea behind this theory is that a large push or a large and comprehensive investment package is necessary to bring economic development and growth. The argument is based on the fact that the size of the domestic market is too small for firms to generate enough sales to make adoption of increasing returns technologies profitable. However, a simultaneous/coordinated investment, such as large infrastructure investment in railways, can generate economic growth if there are strong aggregate demand externalities. Thus, if there exist increasing returns, then a large temporary shock, i.e., investment in local railways, could have a permanent effect, in thus far as it shifts the economy between multiple equilibria.

In this paper, we test two predictions from Murphy et al.'s (1989) model of multiple equilibria based on aggregate demand externalities. Importantly, these predictions differ from other theories of agglomeration economies, i.e., models based on sharing, matching, and learning mechanisms (e.g., Duranton and Puga (2004)).

The first prediction we are going to test is that expectations, and not history (initial condition), determine the choice of equilibrium.[24] The second prediction we test is balanced growth, i.e., all industrial sectors are equally affected by railways.

To test the first prediction of that expectation, and not initial conditions, determine development and economic activity, we divide our panel data set into two groups depending on the initial level of historical economic activity, i.e., the first year when we have reliable data, which is 1871. Thus, we have one group where the level of economic activity is historically low and another group with a high initial level of economic activity. As a result, we can test this prediction of whether the size of the treatment effect differs depending on the initial level of economic activity using our previous event study design, which used railway openings during the period 1880-1917. We expect that the effect of simultaneous/coordinated investment, i.e., local railways in our case, on future local economic activity would be of similar size in both groups if strong aggregate demand externalities exist. On the other hand, if there are mechanisms of agglomeration economies behind our treatment effect other than aggregate demand, then we would expect a much larger effect in the group of local governments with a high initial level of economic activity since this provides support for models of agglomeration economies featuring path dependence.

---

[24] See Krugman (1991) for a discussion of the issues of expectations versus history in determining the selection of equilibrium.

Figures 13 and 14 show the results from the event study for the group of local governments with low and high initial levels of economic activity, respectively. Both of these graphs show very similar evaluations of dynamic treatment effects and no evidence of violation of parallel trends. There is even some evidence that the long-run treatment effect after 30 years is even larger for the group of local governments with an initially low level of economic activity, i.e., 112 log points vs. 74.[25] Thus, the results from these two event study designs provide little evidence that the size of the treatment effect depends on favorable initial conditions. Thus, we find support that expectations rather than history determine the equilibrium outcome.

Figure 13. Event study design: the effect of railways on (log) real nonagricultural income using the sample of local governments with low historical economic activity

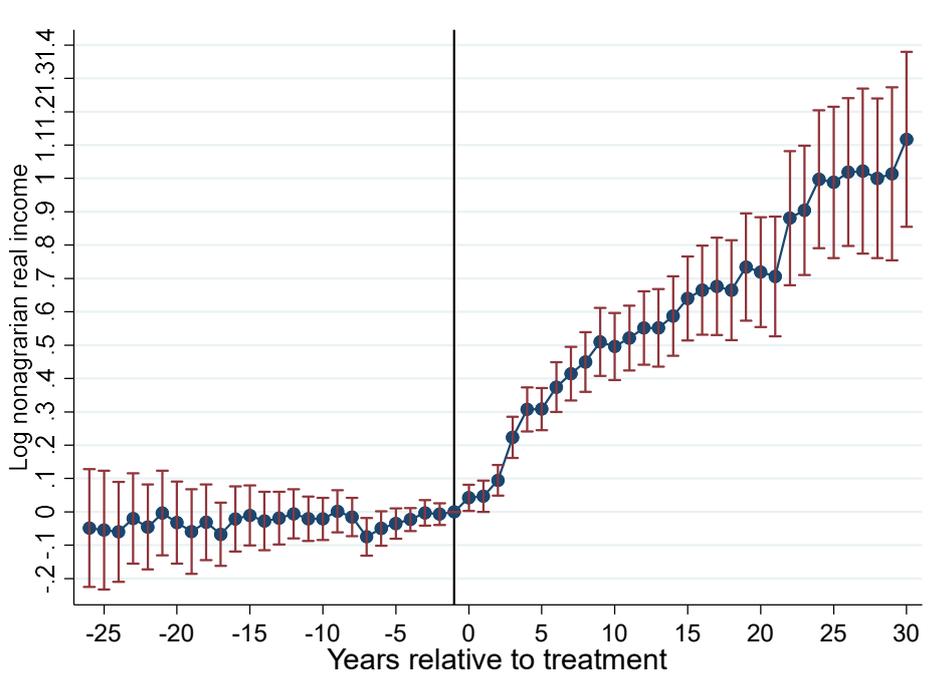

Note: The event study design is implemented using the estimator developed by de Chaisemartin and d'Haultfoeuille (2020, 2021) and the did_multiplegt Stata command.

---

[25] However, it is important to stress that the number of switchers that identifies the long-run effect after 30 years in both these event studies are small, i.e., 47 and 89, respectively.

Figure 14. Event study design: the effect of railways on (log) real nonagricultural income using the sample of local governments with high historical economic activity

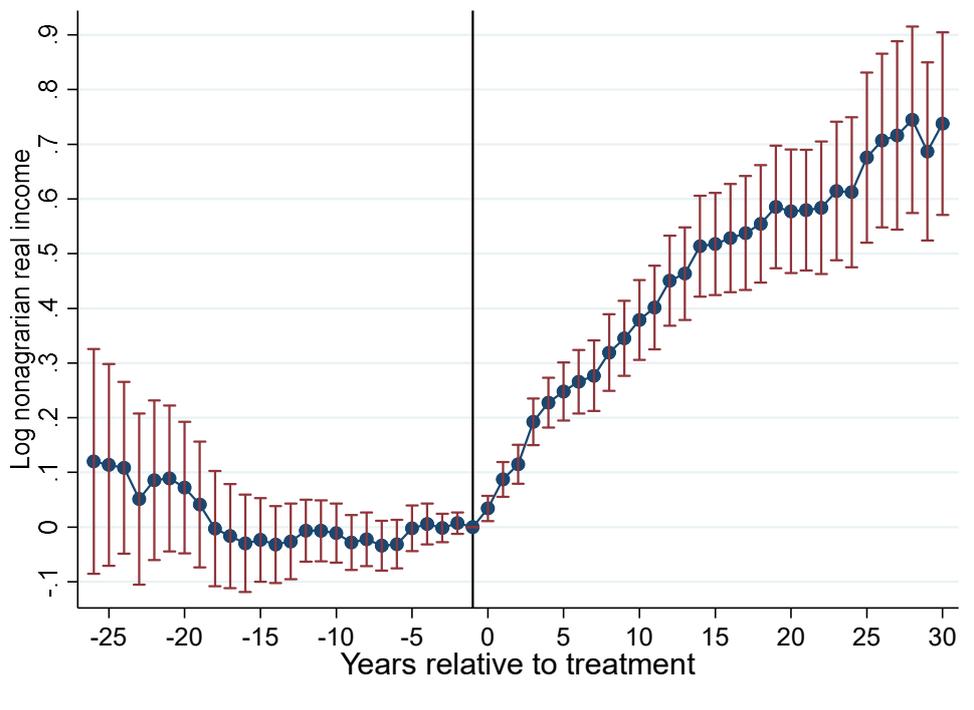

Note: The event study design is implemented using the estimator developed by de Chaisemartin and d'Haultfoeuille (2020, 2021) and the did_multiplegt Stata command.

Turning to the test of the second prediction of balanced growth, i.e., a synchronized expansion across all industries. To perform such a test, we collected and digitized plant (establishment)-level data on the universe of industrial firms included in the official industrial statistics for the years 1913-1917.[26] Importantly, the plant-level data include information about the type of industry to which the plant belongs. Specifically, the data include the following 9 main industrial sectors:

---

[26] Unfortunately, the Swedish manufacturing census before 1913 is unreliable in many respects as discussed by Statistics Sweden in various publications. For example, many of the most important industries are not covered and the plant level statistics are not trustworthy due to issues of double counting. As a result, it is not possible to construct a useable plant level data set over the period of investigation 1860-1917. Indeed, in the publication Historical Statistics in Sweden (1960), they even have separate tables for the industrial statistics for the periods 1836-1895 (Table 5), 1896-1912 (Table 6), 1913-1950 (Table 8) underscoring the problem of comparability across years. See link
http://share.scb.se/OV9993/Data/Historisk%20statistik/Historisk%20statistik%20f%C3%B6r%20Sverige%201700-1900-tal/Statistiska-oversiktsabeller-utover-i-del-I-och-del-II-publicerade-tom-ar-1950.pdf

(1) Ore mining, metal industries, manufacture of metal products

(2) Nonmetallic mining and quarrying and manufacturing

(3) Manufacture of wood and cork

(4) Manufacture of paper and paper products, printing and allied industries

(5) Food manufacturing industries

(6) Manufacture of textiles, wearing apparel and made-up textile goods

(7) Manufacture of leather, furs and rubber products

(8) Electricity, gas and water services

(9) Manufacture of chemicals and chemical products

With these data, we can construct aggregated measures of both production value and employment at the appropriate geographical level where aggregate demand externalities are likely to operate, i.e., at the same level as the treatment is defined (i.e., local infrastructure investments in railways). Additionally, we aggregated the plant-level data over 5 years (i.e., 1913-1917) to mitigate problems with any business cycle issues.

We will perform a simple cross-sectional regression design with a single binary variable, *Rail*, taking the value of one if the local government has a local railway and zero otherwise, i.e.,

(3)   $Y_i = \alpha + \beta Rail_i + \varepsilon_i$

Thus, this cross-sectional regression compares the difference in means between local railways with and without railways at the end of our sample period, i.e., 1917. The aggregate plant-level outcomes we are going study are the sum of the total production value and the sum of total employment of plants located in local government *i*.

Most importantly, we also run a separate regression for all nine major industrial sectors to test for balanced growth, i.e., whether all industrial sectors are affected similarly by the treatment (i.e., local railways). Equally important, we also test whether these results depend on the initial conditions by dividing them into two groups: low vs. high initial economic activity. Thus, there will be a total of 18 subgroups. Another benefit of stratifying the data based on both industry and initial conditions is that subclassification mitigates problems with omitted variable bias, as discussed by Imbens and Rubin (2015: Chapter 17), among others. Again, we would expect that initial conditions do not matter for economic development if expectations, rather than history, determine the choice of equilibrium.

Table 2 shows the results for the (log) value of industrial production, while Table 3 displays the corresponding results for industrial employment. Panel A shows the results for all local governments, while panel B and Panel C display the results for those with a low initial and high level of economic activity, respectively.

Starting with the results from industrial production, Panel An in Column 1 in Table 1 shows that production is increased by 111 log points or 200% (=exp(1.11) -1) if a local government has invested in a local railway. This is a very large treatment effect that is even larger than the results from the event study design. However, the event study can only estimate the long-run effect after 30 years, whereas the cross-sectional design implicitly compares treatment for a much longer time, i.e., from when the first railway was built in 1856 to 1917.[27] Moreover, the estimate from the cross-sectional design is only based on data from 1,342 out of a total of approximately 2,400 rural governments. This fact has to do with industrial statistics placing restrictions on the minimum value of production. Thus, all small industrial firms are not included in industrial statistics. As a result, 56% of all rural governments lack firm data. The treatment group in Panel A in Column 1 therefore includes 893 local governments in the treatment group (those with railways in 1917) and 449 in the comparison group (those without railways in 1917). The treatment group included data from 3.8 plants, on average, on a yearly basis, while the comparison group included only 2.4 plants. Thus, the treatment group included data from 1.4 more plants than the control group.

Turning to the results from the analyses of the main industrial sectors, Columns 2–10 in Panel A report these results for all local governments. Panel A shows that all sectors, except one, increase their production significantly if they have invested in a local railway. The treatment effect ranges from 72 to 165 log points or in the range of 100–320%. Equally important, we also find that the treatment effects are of similar sizes for those local governments with a low level of economic activity (Panel B) and those with a high level (Panel C). Thus, our results in Table 2 show that investments in local railways have a dramatic effect on industrial output in almost all sectors, lending support to the theory of the big push.

Turning to the effect of investments in local railways on industrial employment, we perform exactly the same regression analysis as we did in Table 2. Table 3 presents these results. Column 1 in Panel A shows that employment is increased by 88 log points or 140% (=exp(0.88) -1) if a local government has invested in a railway. Again, this is a very large treatment effect but is smaller than the corresponding effect on production in Table 2. Once again, we find that all industrial sectors except one see a very large increase in industrial employment in the range of 85 to 320%. Moreover, the size of the treatment effect does not seem to differ much between local governments with a low (Panel B) versus a high initial level of economic activity (Panel C). Taken together, the results from Table 3 also lend support to the theory of the big push.

---

[27] Reassuringly, when we restrict our sample based on railway openings during the period 1880-1917, i.e., the same period as in the event study, the estimated treatment effect is 0.85. Thus, this effect is nearly of the same magnitude as in the event study analysis of 0.80.

Table 2. The effect of railways on the (log) industrial production value

| | All sectors | Sector 1 | Sector 2 | Sector 3 | Sector 4 | Sector 5 | Sector 6 | Sector 7 | Sector 8 | Sector 9 |
|---|---|---|---|---|---|---|---|---|---|---|
| | (1) | (2) | (3) | (4) | (5) | (6) | (7) | (8) | (9) | (10) |
| **Panel A. All local governments** | | | | | | | | | | |
| Railway | 1.11 | 1.22 | 0.97 | 1.15 | 0.78 | 1.18 | 0.72 | 0.79 | 1.65 | -0.32 |
| | (0.11) | (0.26) | (0.20) | (0.15) | (0.26) | (0.17) | (0.46) | (0.52) | (0.46) | (0.46) |
| **Panel B. Local governments with a low initial level of economic activity** | | | | | | | | | | |
| Railway | 0.94 | 1.32 | 0.76 | 1.09 | 0.78 | 1.05 | 0.50 | 1.77 | 1.94 | -0.41 |
| | (0.13) | (0.42) | (0.25) | (0.18) | (0.46) | (0.23) | (0.79) | (0.59) | (0.54) | (0.97) |
| **Panel C. Local governments with a high initial level of economic activity** | | | | | | | | | | |
| Railway | 1.08 | 1.17 | 0.85 | 0.93 | 0.55 | 1.08 | 0.80 | 0.48 | 1.54 | -0.29 |
| | (0.17) | (0.31) | (0.29) | (0.21) | (0.30) | (0.24) | (0.51) | (0.64) | (0.80) | (0.47) |
| Obs. in Panel A | 1,342 | 434 | 430 | 971 | 165 | 678 | 112 | 83 | 94 | 184 |
| Obs. in Panel B | 648 | 117 | 200 | 332 | 32 | 266 | 35 | 19 | 25 | 42 |
| Obs. in Panel C | 694 | 317 | 230 | 459 | 133 | 412 | 77 | 64 | 69 | 142 |

Notes: Each entry is a separate regression. Robust standard errors are within parentheses.

Table 3. The effect of railways on (log) industrial employment

| | All sectors | Sector 1 | Sector 2 | Sector 3 | Sector 4 | Sector 5 | Sector 6 | Sector 7 | Sector 8 | Sector 9 |
|---|---|---|---|---|---|---|---|---|---|---|
| | (1) | (2) | (3) | (4) | (5) | (6) | (7) | (8) | (9) | (10) |
| | Panel A. All local governments | | | | | | | | | |
| Railway | 0.88 | 1.05 | 0.61 | 0.95 | 0.69 | 0.99 | 0.76 | 1.02 | 1.44 | -0.27 |
| | (0.10) | (0.23) | (0.15) | (0.13) | (0.22) | (0.16) | (0.40) | (0.47) | (0.43) | (0.40) |
| | Panel B. Local governments with a low initial level of economic activity | | | | | | | | | |
| Railway | 0.95 | 1.38 | 0.52 | 0.96 | 0.83 | 0.91 | 0.53 | 2.27 | 2.02 | -0.90 |
| | (0.13) | (0.40) | (0.18) | (0.15) | (0.40) | (0.19) | (0.61) | (0.54) | (0.76) | (0.71) |
| | Panel C. Local governments with a high initial level of economic activity | | | | | | | | | |
| Railway | 1.09 | 0.89 | 0.46 | 0.72 | 0.42 | 0.84 | 0.85 | 0.63 | 1.24 | -0.04 |
| | (0.17) | (0.27) | (0.22) | (0.18) | (0.24) | (0.21) | (0.49) | (0.55) | (0.46) | (0.46) |
| Obs. in Panel A | 1,343 | 435 | 430 | 971 | 165 | 678 | 112 | 83 | 94 | 184 |
| Obs. in Panel B | 648 | 118 | 200 | 332 | 32 | 266 | 35 | 19 | 25 | 42 |
| Obs. in Panel C | 694 | 317 | 230 | 459 | 133 | 412 | 77 | 64 | 69 | 142 |

Notes: Each entry is a separate regression. Robust standard errors are within parentheses.

# 8. Conclusion

In this paper, we analyzed the effect of transport infrastructure investments in railways on key measures of local economic activity, such as real nonagricultural income, agricultural land values, population size, industrial employment and production.

As a testing ground, we used data from a new historical database that includes *annual* panel data on approximately 2,400 rural regions (i.e., rural local governments during the period 1860–1917). We used a staggered event study design that is robust to treatment effect heterogeneity, i.e., de Chaisemartin and d'Haultfoeuille (2020, 2021). Importantly, we find extremely large reduced-form effects of having access to railways. For real nonagricultural income, the cumulative treatment effect is approximately 120% after 30 years. Therefore, this effect is 20 times larger than most of the reduced-form effects found in previous work on the effect of transport infrastructure on economic activity (Redding and Turner 2015). Equally important, we also show that our reduced-form effect probably reflects growth rather than a reorganization of existing economic activity since we find no spillover effects between treated and untreated regions. Thus, the absence of spillover effects also means that the stable unit treatment value assumption (SUTVA) holds in our setting.

We also provide suggestive evidence on the mechanism that gives rise to such a substantial effect of investments in local railways on local economic activity by using plant-level data. Indeed, we show that both local industrial production and employment are 100–300% larger across almost industrial sectors in rural areas connected by local railways than in areas without railways. These results also hold in areas with both a historically low and high level of economic activity. As a result, these results are consistent with a big push hypothesis, i.e., that a simultaneous/coordinated investment (a big push), such as large infrastructure investment in railways, can generate economic growth, independent of favorable initial conditions, if there are strong aggregate demand externalities,

# References

Acemoglu, Daron, Simon Johnson, and James A. Robinson. (2005). "Institutions as a Fundamental Cause of Long-run Growth." Handbook of Economic Growth, 1, 385-472.

Andersson, David, Thor Berger and Erik Prawitz (2020), "Making a Market: Infrastructure, Integration and the Rise of Innovation," forthcoming at Review of Economics and Statistics.

Andersson-Skog, Lena (1993). "Såsom allmänna inrättningar till gagnet, men affärsföretag till namnet: SJ, järnvägspolitiken och den ekonomiska omvandlingen efter 1920." Doctoral dissertation, Umeå universitet.

Atack, Jeremy and RA. Margo (2011). "The impact of access to rail transportation on agricultural improvement: the American Midwest as a test case." J. Transp. Land Use 4:5–18.

Banerjee, Abhijit Esther Duflo and Nancy Qian (2020) "On the Road: Transportation Infrastructure and Economic Growth in China" forthcoming at The Journal of Development Economics.

Berger T. (2019). "Railroads and Rural Industrialization: evidence from a Historical Policy Experiment," Explorations in Economic History, Volume 74,

Berger, T., and Enflo, K. (2017). "Locomotives of local growth: The short-and long-term impact of railroads in Sweden." *Journal of Urban Economics*, *98*, 124-138.

Bleakley, H., and J. Lin (2012). "Portage and Path Dependence," The Quarterly Journal of Economics, 127(2), p. 587-611.

Bogart, Dan (2020). "Infrastructure and institutions: Lessons from history," forthcoming in Regional Science and Urban Economics

Borusyak, Kirill, Xavier Jaravel and ann Spiess (2021). "Revisiting Event Study Designs: Robust and Efficient Estimation." mimeo

Callaway, Brantly, and Pedro H.C. Sant'Anna, (2020). "Difference-in-Differences with multiple time periods," Journal of Econometrics,

de Chaisemartin, Clement, and Xavier d'Haultfoeuille (2020). "Two-way fixed effects estimators with heterogeneous treatment effects." American Economic Review 110.9: 2964-96.

de Chaisemartin, Clement, and Xavier d'Haultfoeuille (2021). "Difference-in-Differences Estimators of Intertemporal Treatment Effects." Working paper

Davis, D., and D. Weinstein (2002). "Bones, Bombs, and Break Points: The Geography of Economic Activity." American Economic Review, 2(5), 1269-1289.

Donaldson, Dave (2015). "The Gains from Market Integration." Annual Review of Economics. Vol. 7:619-647

Donaldson, Dave (2018). "Railroads of the Raj: Estimating the Impact of Transportation Infrastructure." American Economic Review 2018, 108(4-5): 899–934.

Donaldson, Dave and Richard Hornbeck (2016). Railroads and American Economic Growth: A "Market Access" Approach, The Quarterly Journal of Economics, Volume 131, Issue 2, pages 799–858.

Duranton, Gilles and Diego Puga (2004) "Chapter 48 - Micro-Foundations of Urban Agglomeration Economies," Editor(s): J. Vernon Henderson, Jacques-François Thisse, Handbook of Regional and Urban Economics, Elsevier, Volume 4, Pages 2063-2117.

Duranton, G., & Kerr, W. R. (2015). The logic of agglomeration (No. w21452). National Bureau of Economic Research.

Flodström, I. (1912). Sveriges nationalförmögenhet omkring år 1908 och dess utveckling sedan midten av 1880-talet. Finansstatistiska utredningar 5. Stockholm: Finansdepartementet.

Fogel, R. W. (1964). Railroads and American economic growth: Essays in econometric history. Baltimore: Johns Hopkins Press.

Goodman-Bacon, Andrew (2021). "Difference-in-Differences with Variation in Treatment Timing," Journal of Econometrics

Heblich, Stephan, Stephen J Redding, Daniel M Sturm (2020), "The Making of the Modern Metropolis: Evidence from London," *The Quarterly Journal of Economics*, Volume 135, Issue 4, pages 2059–2133,

Heckscher, E. F. (1907). Till belysning af järnvägarnas betydelse för Sveriges ekonomiska utveckling. Centraltryckeriet. Stockholm.

Hedin, L E (1967), Some notes on the financing of the Swedish railroads. Economy and History. Vol X.

Holgersson, B., and Nicander, E. (1968). "The Railroads and the Economic Development in Sweden during the 1870s." Economy and History, 11(1), 3-51.

Hornung, Erik, (2015). "Railroads and Growth in Prussia," Journal of the European Economic Association, Volume 13, Issue 4, 1, Pages 699–736

Imbens, G. W., & Rubin, D. B. (2015). Causal inference in statistics, social, and biomedical sciences. Cambridge University Press.

Johansson, D., Stenkula, M. och Du Rietz, G. (2015). "Capital Income Taxation of Swedish Households, 1862 to 2010", Scandinavian Economic History Review, vol. 63:2, 154–177.

Kline, Patrick and Enrico Moretti (2014). "Local Economic Development, Agglomeration Economies, and the Big Push: 100 Years of Evidence from the Tennessee Valley Authority," The Quarterly Journal of Economics, Volume 129, Issue 1, Pages 275–331

Krugman, P. (1991a), "History versus Expectations", Quarterly Journal of Economics 106, 651–667.

Lin, Jeffrey and Ferdinand Rauch (2020) "What Future for History Dependence in Spatial Economics?", forthcoming in Regional Science and Urban Economics.

Lindgren, Erik, Per Pettersson-Lidbom and Björn Tyrefors (2019). "The Political Economics of Growth, Labor Control and Coercion: Evidence from a Suffrage Reform." IFN Working Paper nr 1172.

Lindgren, Erik, Per Pettersson-Lidbom and Björn Tyrefors (2021). "The Causal Effect of Political Power on the Provision of Public Education: Evidence from a Weighted Voting System" IFN Working Paper nr 1315.

Lindgren, Håkan (2017) "Om fastighetsvärderingar i svenska bouppteckningar under 1800 talet." Working paper presented at the 12th economic history meeting in Stockholm.

Nicander, Eric (1980). Järnvägsinvesteringar i Sverige 1849-1914 (Vol. 28). Ekonomisk-historiska föreningen.

Mellquist, Einar (1974). Rösträtt efter Förjänst (Voting According to Income? The Parliamentary Debate about the Municipal Voting System in Sweden 1862-1900). Ph. D. Dissertation, Uppsala University

Matsuyama, Kiminori (1991). "Increasing Returns, Industrialization, and Indeterminacy of Equilibrium," The Quarterly Journal of Economics, Volume 106, Issue 2, Pages 617–650

Modig, Hans (1971), "Järnvägarnas efterfrågan och den svenska industrin 1860-1914", Ph. D. Dissertation Uppsala University.

Murphy, K., Shleifer, A., & Vishny, R. (1989). "Industrialization and the Big Push." Journal of Political Economy, 97(5), 1003-1026

Oredsson, Sverker (1969). Järnvägarna och det allmänna. Gleerupa.

Oredsson, Sverker (1989) "Järnvägsbyggandet och den kommunala kompetensen." Historisk Tidskrift.

Ottosson, Jan and Lena Andersson-Skog (2013). "*Stat, marknad och reglering i historiskt perspektiv*", Konkurrensverkets uppdragsforskningsrapport 2013:3, Konkurrensverket. https://www.konkurrensverket.se/globalassets/aktuellt/nyheter/stat-marknad-och-reglering-i-historisk-perspektiv.pdf

Redding, Steven (2020) "Trade and Geography", NBER Working Paper, 27821.

Redding, S. J., and Turner, M. A. (2015). Transportation costs and the spatial organization of economic activity. *Handbook of Regional and Urban Economics*, 5, 1339-1398.

Rodríguez-Clare, A. (1997). "Positive Feedback Mechanism in Economic Development: A Review of Recent Contributions." in Development Strategy and Management of the Market Economy, I. P. Szekeley and R. Sabot (eds), Clarendon Press Oxford, 1997.

Rosenstein-Rodan, P. N., (1943). "Problems of Industrialisation of Eastern and South-Eastern Europe," The Economic Journal, Volume 53, Issue 210-211, Pages 202–211.

Rosenthal, S. S., & Strange, W. C. (2020). How close is close? The spatial reach of agglomeration economies. Journal of economic perspectives, 34(3), 27-49.

Stenkula, M. (2014). "Swedish Taxation in a 150-year Perspective". Nordic Tax Journal, vol. 1:2, 10–42.

Sun, Liyang, and Sarah Abraham (2020). "Estimating dynamic treatment effects in event studies with heterogeneous treatment effects." Journal of Econometrics.

# Appendix

In this appendix, we empirically illustrate the problem of using a standard two-way fixed effect (TWFE) estimator to estimate the effects of railways on economic activity when the treatment effect is likely to be heterogeneous. We also compare the results from the TWFE approach with an estimator that is robust to treatment heterogeneity. The illustration is based on four recent and prominent studies with highly accessible data: Donaldson (2018), Heblich et al. (2020), Hornung (2015), and Donaldson and Hornbeck (2016). Moreover, we also test whether the TWFE estimator produces biased results in our study.

There are at least three problems with a TWFE estimator when treatment effects are heterogeneous according to Borusyak et al. (2021), i.e., underidentification of the fully dynamic regression, negative weights in both the static and dynamic regressions, and spurious identification of the long-term effects. As will become clear below, all four of the studies had at least one of these problems. As a result, we show that the estimates from their TWFE specifications are not reliable.

Starting with Donaldson (2018), he used yearly data from 235 districts in India over the period of 1870-1930.[28] He used a static TWFE specification, i.e.,

(1) $$Y_{it}=\alpha_i + \lambda_t + \beta X_{it} + \varepsilon_{it},$$

where $i$ denotes a unit, and $t$ denotes time. $Y_{it}$ is the log of real income in the agricultural sector, and $X_{it}$ is a binary variable taking the value of 1 if the district has access to a railway in period $t$.

He reported a point estimate of $\beta$ as 0.164 with a cluster robust standard error of 0.049. Thus, according to his TWFE estimate, access to a railway increases real income in the agricultural sector by 18% (=exp(0.164)-1). However, the static TWFE specification places severe restrictions on the dynamic effects of the treatment, leading to bias unless they are true. To avoid this type of bias, we can allow for an unrestricted (saturated) dynamic specification by respecifying equation (1) as an event study, i.e., a TWFE dynamic event study regression, as follows:

(2) $$Y_{it}=\alpha_i + \lambda_t + \sum_{k=-K}^{-2} \beta_k^{lead} X_{it}^k + \sum_{k=0}^{L} \beta_k^{lag} X_{it}^k + \varepsilon_{it},$$

---

[28] However, there are a very large number of missing observations, making the panel highly unbalanced.

where $X_{it}^k$ is an event-study indicator variable that takes the value of one if unit *i* is *k* periods away from the initial treatment at time *t* and zero otherwise. Figure 1A shows the results from the event study with an unrestricted dynamic specification, where the first lead is excluded for normalization.[29] Importantly, there is no problem with underidentification in the unrestricted dynamic specification due to the existence of never-treated units in Donaldson's data.

For clarity, we only present the estimates for the 10 years before and after the treatment occurs. The results from the event study design show that we cannot reject that the parallel trend assumption holds but also that we cannot reject that there is no effect of the treatment. Thus, the dynamic TWFE specification yields very different results from the static TWFE specification used by Donaldson.

Figure 1A. Results from a TWFE specification with unrestricted dynamics: Donaldson (2018)

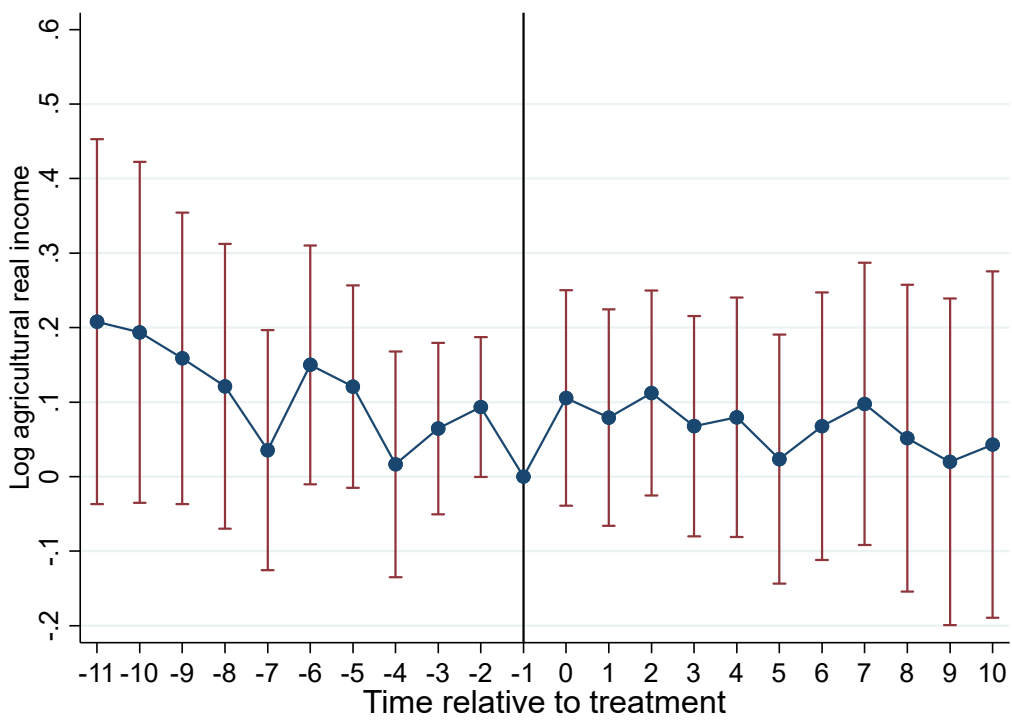

To further investigate whether the dynamic TWFE specification is reliable,[30] we also estimate the event study design using the de Chaisemartin and D'Haultfoeuille (2020, 2021) estimator, which is robust to heterogeneous treatment effects. This result is presented in Figure 2A and reveals a similar result to the fully dynamic TWFE specification.

---

[29] This figure is produced using the eventdd command developed by Clarke and Schythe (2020). We used this command to produce other fully dynamic event study specifications in this Appendix.
[30] Sun and Abraham (2021) showed that the unrestricted dynamic TWFE specification could be unreliable if the treatment effect was heterogenous.

Figure 2A. Results from the de Chaisemartin and D'Haultfoeuille estimator: Donaldson (2018)

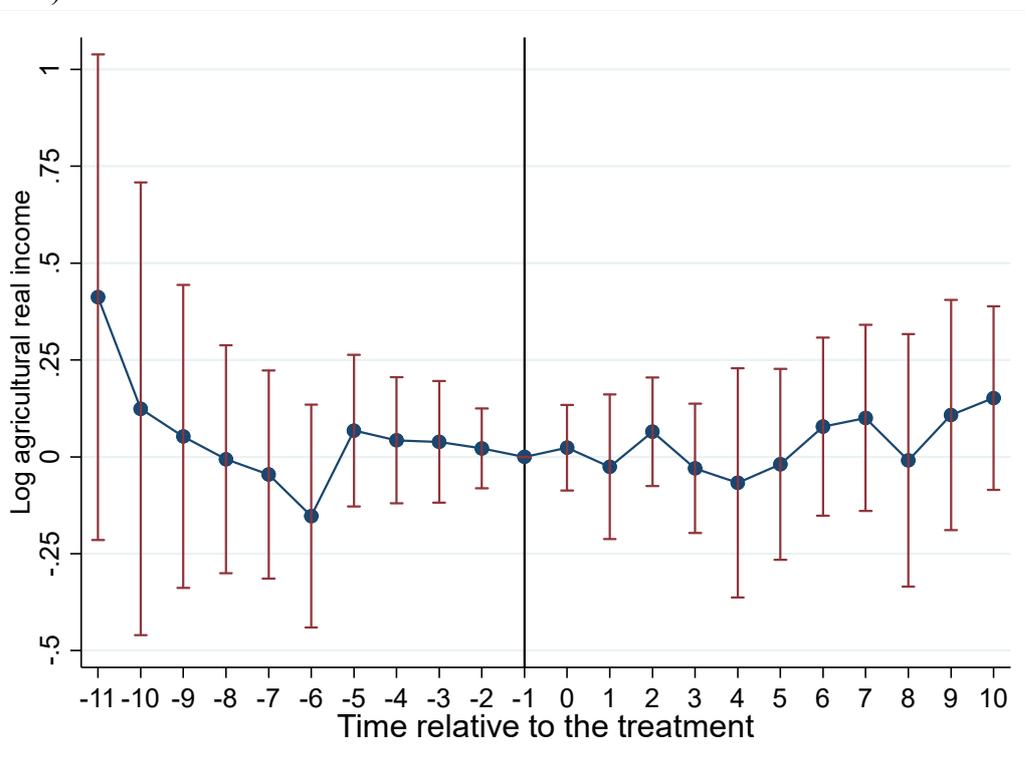

Heblich et al. (2020) analyzed the effects of railways/subways on log population size using parish-level data from London on a decadal basis over the period of 1801-1901.[31] The number of parishes in their data is 283, and 128 of them obtained access to a railway/subway during this period.

Heblich et al. used a dynamic TWFE specification but placed restrictions on the fully dynamic specification, i.e., they binned their endpoints. However, there is no reason to restrict the dynamic specification because of the existence of a large group of never-treated units, i.e., 155, in their data. They also included unit-specific linear time trends in their dynamic TWFE specification.

Figure 3A shows the estimates from their specification. This reveals that the parallel trend assumption is likely to hold and that the effect of having access to railways is approximately 10%. The long-term cumulative dynamic effect after 60 years is also substantial, i.e., approximately 320% (=exp(1.43) -1). However, since they restricted their dynamic TWFE specification, it might have introduced bias. Figure 4A shows the results from their dynamic TWFE specification without any restrictions on the dynamic event study specification. Figure 4A shows completely different results from the event study in Figure 3A. Indeed, the parallel trend assumption is violated, and there is no evidence of any dynamic treatment effects. Thus, the TWFE approach yields very different results depending on the dynamic specification.

---

[31] Another issue with their event study is that their data were not annual but only decadal. Thus, this difference means that their event study design relied on observations that were not necessarily close to the event date, i.e., the year of opening of a new subway line. Moreover, there would also be a mismatch between when the actual event took place and when it was recorded. For example, if the event occurred in 1882, it was recorded in the data as having taking place in 1891.

Figure 3A. Results from a restricted event study specification: Heblich et al. (2020)

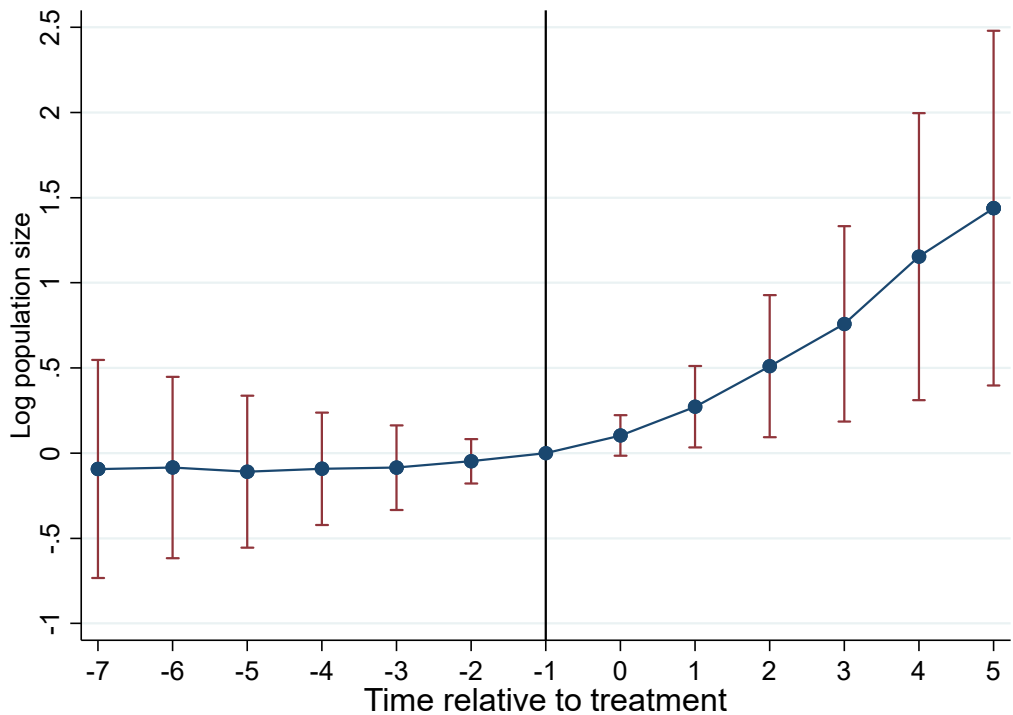

Figure 4A. Results from a TWFE specification with unrestricted dynamics: Heblich et al. (2020)

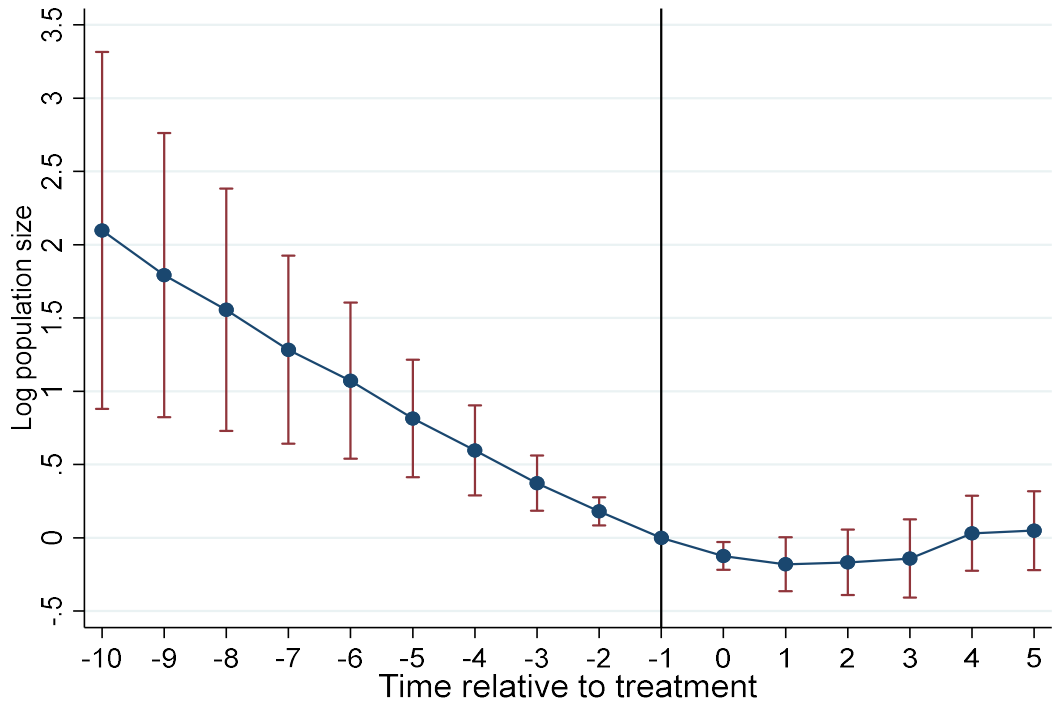

To further probe whether the unrestricted dynamic TWFE specification is reliable, we again estimate the event study design using the de Chaisemartin and D'Haultfoeuille (2020, 2021) estimator, which is robust to heterogeneous treatment effects. This result is presented in Figure 5A. This figure also shows that the parallel trend assumption does not hold. Indeed, there is a strong pretrend before the treatment occurs, casting doubt on the appropriateness of using an event study design in this case.

Figure 5A. Results from the de Chaisemartin and D'Haultfoeuille estimator: Heblich et al. (2020)

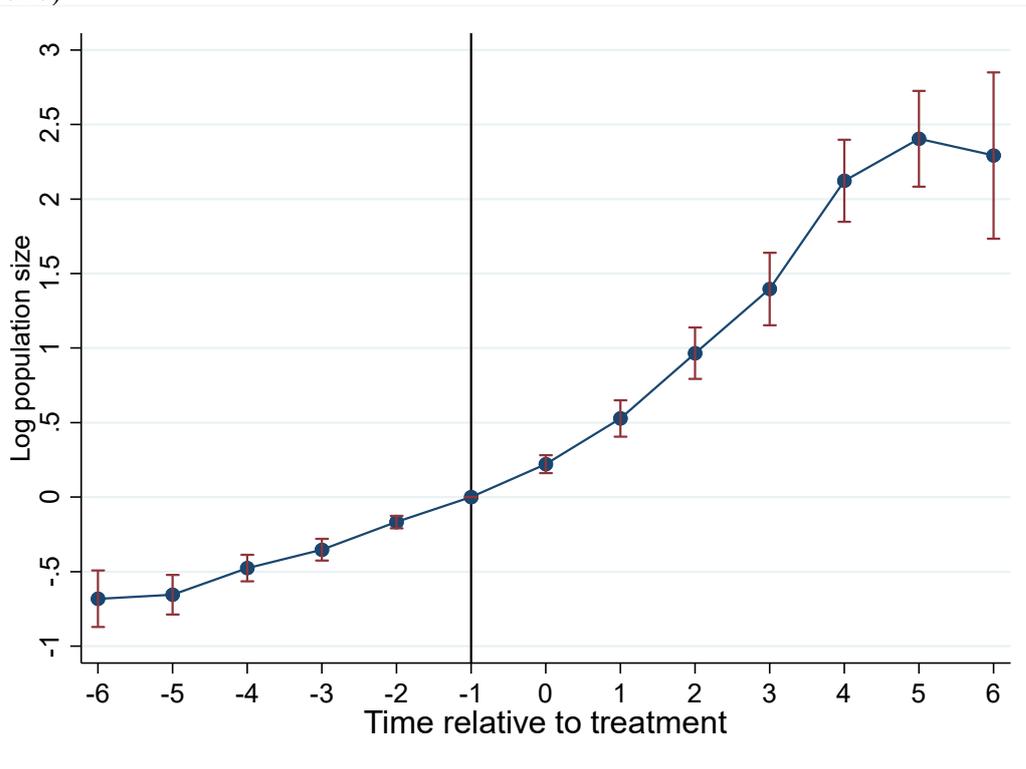

Turning to the study by Hornung (2015), he estimated the effects of railways on log population size using city-level data from the historical German state of Prussia for the period of 1837-1861. His data are not based on annual observations but instead triennial on data from 978 Prussian cities.

Horning also used an event study design with restrictions on the dynamic TWFE specification.[32] Again, there is no reason to place restrictions on the dynamic specification because of the existence of a large group of never-treated units, i.e., 767, in his data. Figure 6A shows the results from his restricted dynamic TWFE specification, demonstrating that the units have parallel trends before the treatment and that the impact effect is approximately 5%. The long-term cumulative treatment effect is approximately 17% (=exp(0.16)-1). However, a restricted dynamic TWFE specification might introduce bias. Figure 7A shows the results from the fully dynamic TWFE specification. Figure 7A reveals that there is a significant pretrend. Thus, once again, the TWFE approach yields different results depending on the dynamic specification.

---

[32] Hornung also defined his leads wrongly in his analysis since he used data after 1861 to define his pretreatment indicator variables, preaccess 2-5. For example, a district receiving access to railways in 1880 would be part of the leads (i.e., preaccess 4 and 5). The correct definition shows that there are significant pretrends using exactly the same dynamic specification as Hornung.

Figure 6A. Results from a restricted event study specification: Hornung (2015)

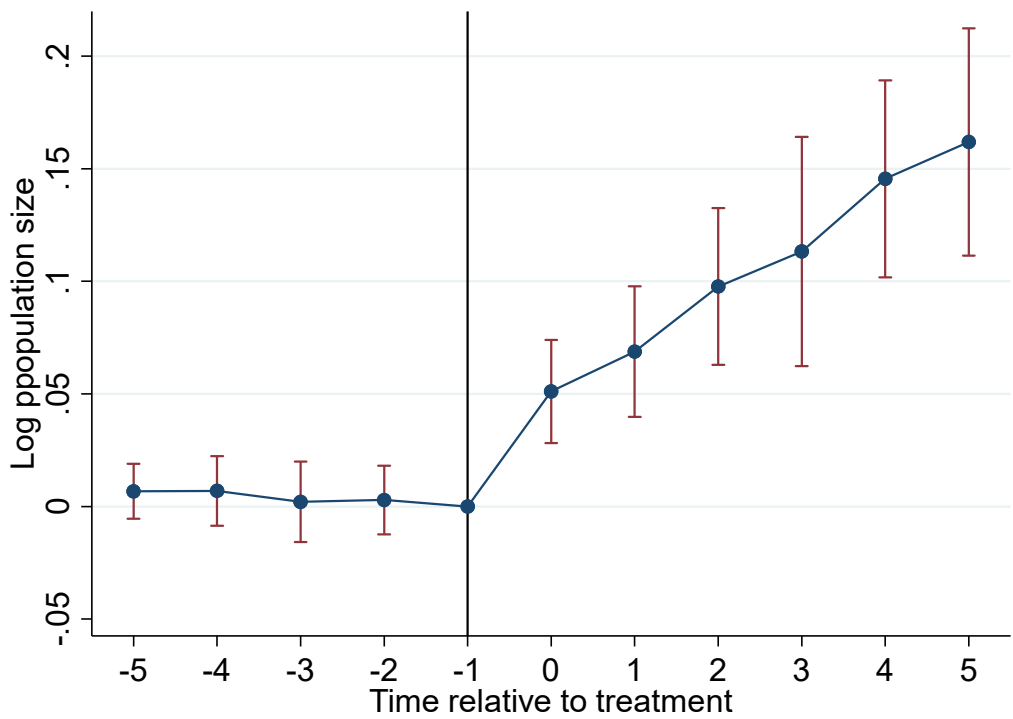

Figure 7A. Results from a TWFE specification with unrestricted dynamics: Hornung (2015)

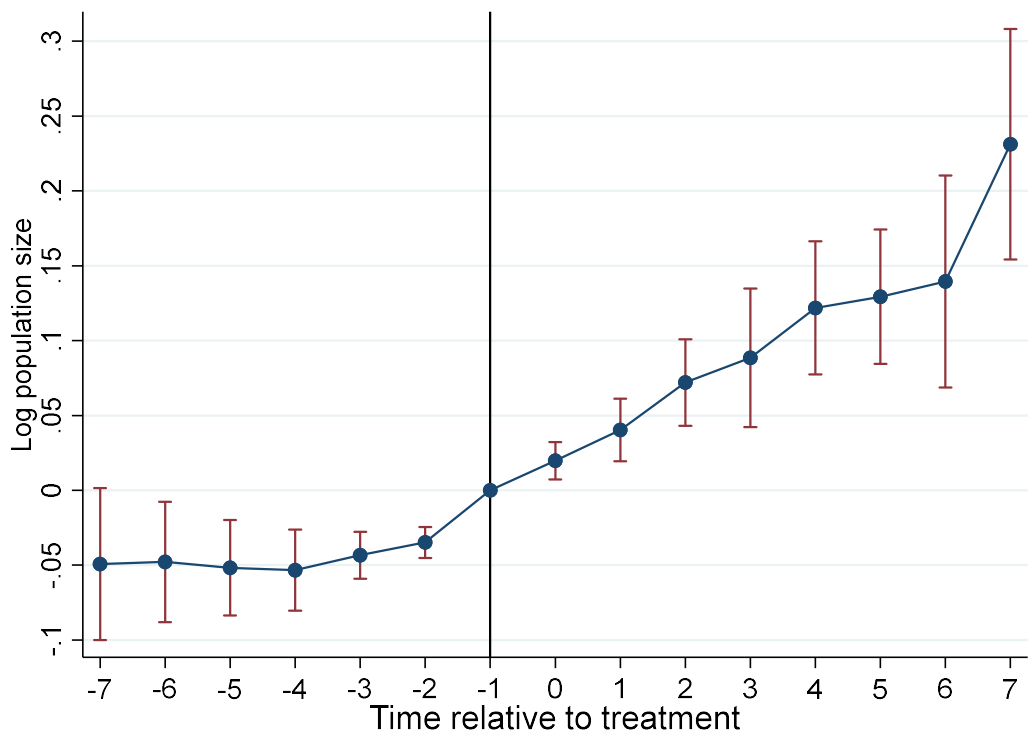

To further probe whether the unrestricted dynamic TWFE specification is reliable, we again estimate the event study design using the de Chaisemartin and D'Haultfoeuille (2020, 2021) estimator, which is robust to heterogeneous treatment effects. This result is presented in Figure 8A. This figure also shows that the parallel trend assumption does not hold. Indeed, there is a strong pretrend before the treatment occurs, again casting doubt on the appropriateness of using an event study design in this case.

Figure 8A. Results from the de Chaisemartin and D'Haultfoeuille estimator: Hornung (2015)

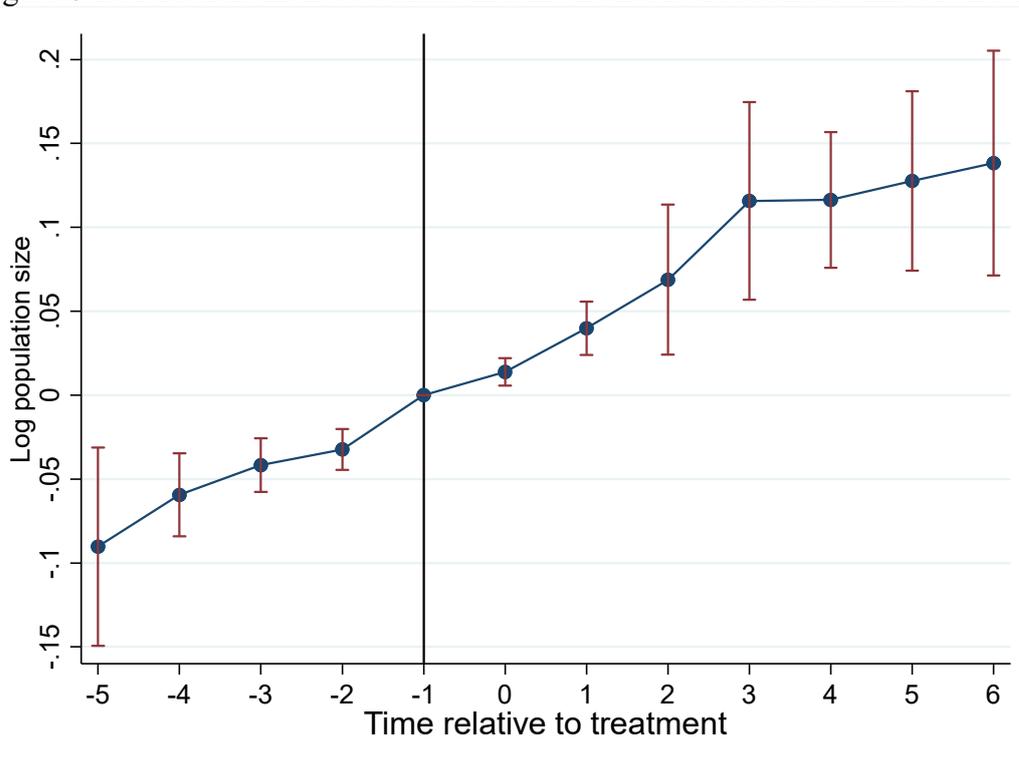

Turning to the study by Donaldson and Hornbeck (2016), this study is different from the previous ones since they estimated a TWFE specification with a continuous variable, i.e., market access, using data from only two time periods, i.e., 1870 and 1890.[33]

However, it is still productive to use their data to test whether a standard TWFE with a binary variable is affected by treatment heterogeneity even if there are only two time periods. In fact, Donaldson and Hornbeck reported such a specification in their NBER working paper.

---

[33] When the treatment variable is continuous, then the treatment effect must be proportional to the observed treatment intensity, as discussed by Schmidheiny and Siegloch (2020). Donaldson and Hornbeck (2016) showed results in their Figure IV suggesting this linearity assumption is fulfilled. However, their analysis was not correct since their residualization approach was inconsistent for the target function of interest, as discussed by Cattaneo et al. (2019). Using the correct residualization approach suggests that the assumption of linearity can be violated.

Specifically, column 3 of table 2 in their working paper reports that county land value increases by 0.290 log points or 34% when a county receives any railroad track, and they argued that "these estimates may reflect a causal impact of railroads on county land values".[34] We can now test whether their TWFE estimator is biased due to dynamic treatment effects by excluding the always treated, i.e., those districts that had a railway in both 1870 and 1890, from the control group, i.e., those districts that did not have a railway in 1870 and 1890. In this case, the point estimate is -0.06 with a standard error of 0.05. Thus, omitting the already treated districts suggests that there is no effect of railways on the outcome. Moreover, the standard errors are so small that one can reject that the two estimates, 0.29 vs. -0.06, are equal. Thus, the TWFE estimator again produces very different results when treatment effect heterogeneity is considered.

In summary, we have analyzed a number of studies, i.e., Donaldson (2018), Heblich et al (2020), Hornung (2015), and Donaldson and Hornbeck (2016), all of which attempted to estimate the causal effects of railways on economic activity using TWFE estimators. We find that the results are sensitive to the dynamic specification of the treatment effect, i.e., restricted versus unrestricted. When we use the unrestricted specification, we often find strong evidence that the parallel trend assumption is violated. When we estimate the effect using an estimator that is robust to treatment effect heterogeneity, we also find similar results of significant pretrends. These results therefore cast doubt on whether these studies have estimated a causal effect of the effects of railways on economic activity.

It might also be interesting to determine whether our results would be different if we used a TWFE estimator. We therefore present estimates from unrestricted dynamic TWFE specifications. To facilitate comparison, we also show the figures from the estimator by de Chaisemartin and D'Haultfoeuille (2020, 2021) that we used in our paper.

Starting with the effects of railways on real income in the nonagricultural sector, Figure 9A shows the results for the fully dynamic TWFE specification, while Figure 10A shows the corresponding results from the de Chaisemartin and D'Haultfoeuille estimators. The results are highly similar. This finding strongly suggests that there is very limited heterogeneity in the treatment effects in our setting.

Continuing with the effect of railways on agricultural land values, Figure 11A shows the results for the fully dynamic TWFE specification, while Figure 12A shows the results from the de Chaisemartin and D'Haultfoeuille estimators. Again, the results are similar except that some of the estimates on the longer leads are now marginally significantly different from zero.

---

[34] Turner and Redding (2015) also interpreted this estimate as representing a causal effect.

Moving on to the effect of railways on population size, Figure 13A shows the results for the fully dynamic TWFE specification, while Figure 14A shows the results from the de Chaisemartin and D'Haultfoeuille estimators. While the estimates of the dynamic effects are similar, the dynamic TWFE estimator now suggests that there is a significant pretrend.[35]

In summary, in our setting, there seem to be small differences in the results between a fully dynamic TWFE estimator and an estimator robust to treatment heterogeneity. It might therefore be interesting to speculate why this finding occurs. There are a number of potential factors that are different in our setting compared to the other studies: (i) the units of observations are very homogenous, which limits the problem of heterogeneity; (ii) we have annual data, which allow us to accurately capture the dynamic treatment response; (iii) we have a very large group of never treated units, which avoids the problem of underidentification; (iv) we have a very large number of uniform events, i.e., openings of primarily local railways, and the events are distributed fairly evenly across time, which helps to mitigate problems with treatment heterogeneity across units and time; and (v) there are no spillover effects, which excludes violations of SUTVA.

---

[35] This result could be due to our having much more data for population size, i.e., 1860-1917, than for the other two outcomes, for which we only have data for the shorter period of 1880-1917. Thus, there will probably be more heterogeneity in the larger sample, leading to bias in the dynamic event study design, as discussed by Sun and Abraham (2020).

Figure 9A. Results from a TWFE specification with unrestricted dynamics: log nonagrarian real income

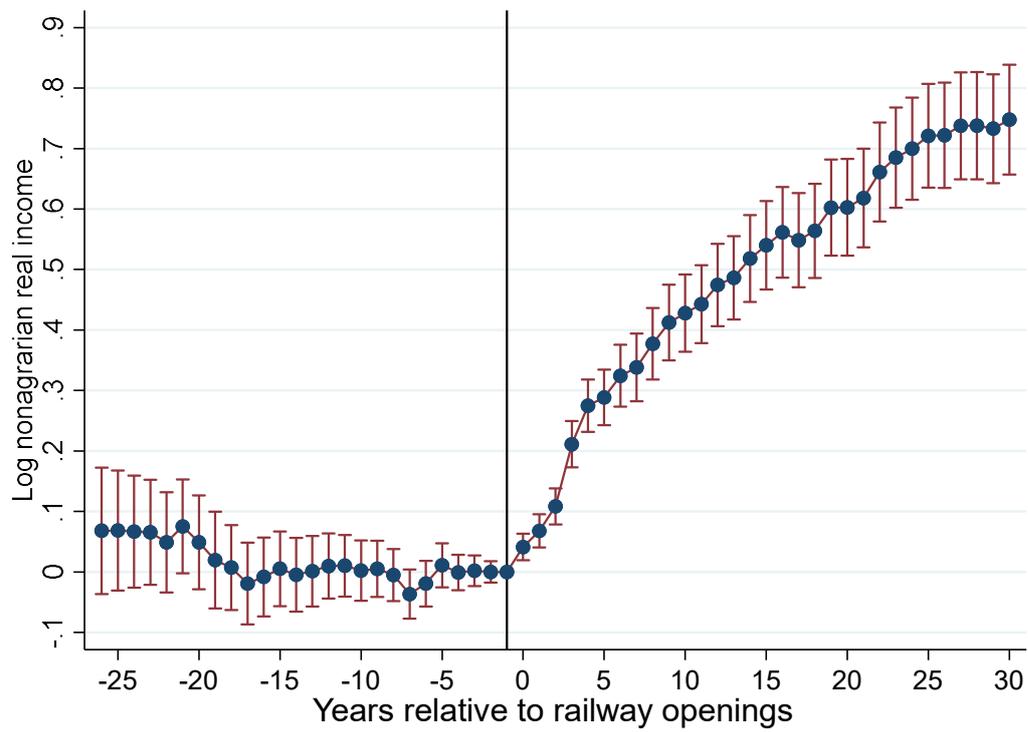

Figure 10A. Results from the de Chaisemartin and D'Haultfoeuille estimator: log nonagrarian real income

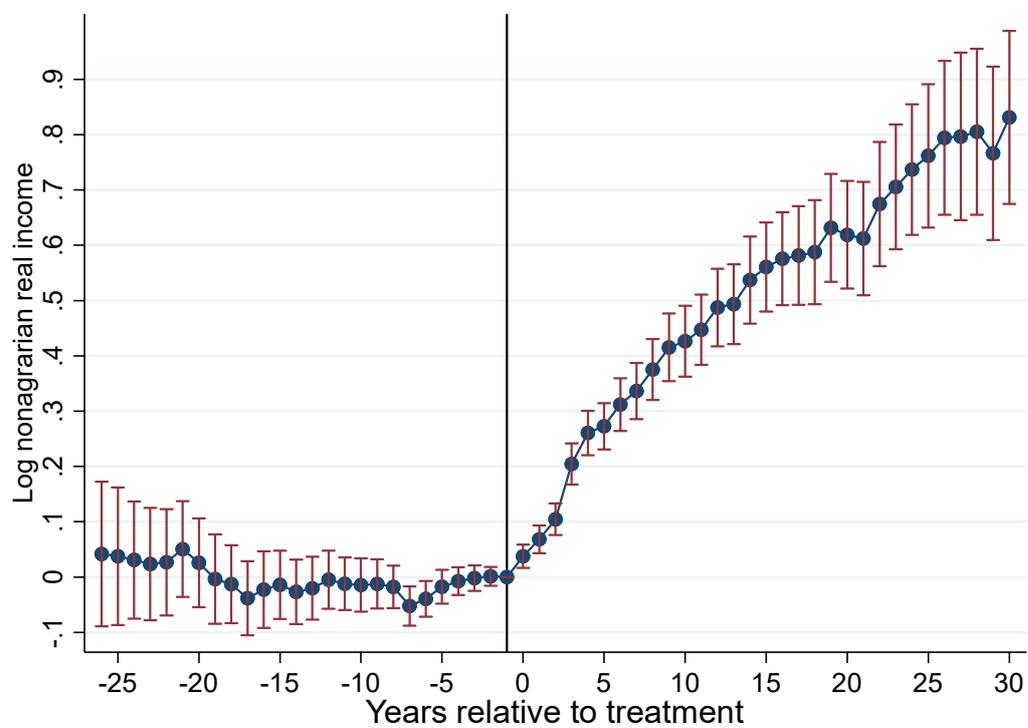

Figure 11A. Results from a TWFE specification with unrestricted dynamics: agricultural land values

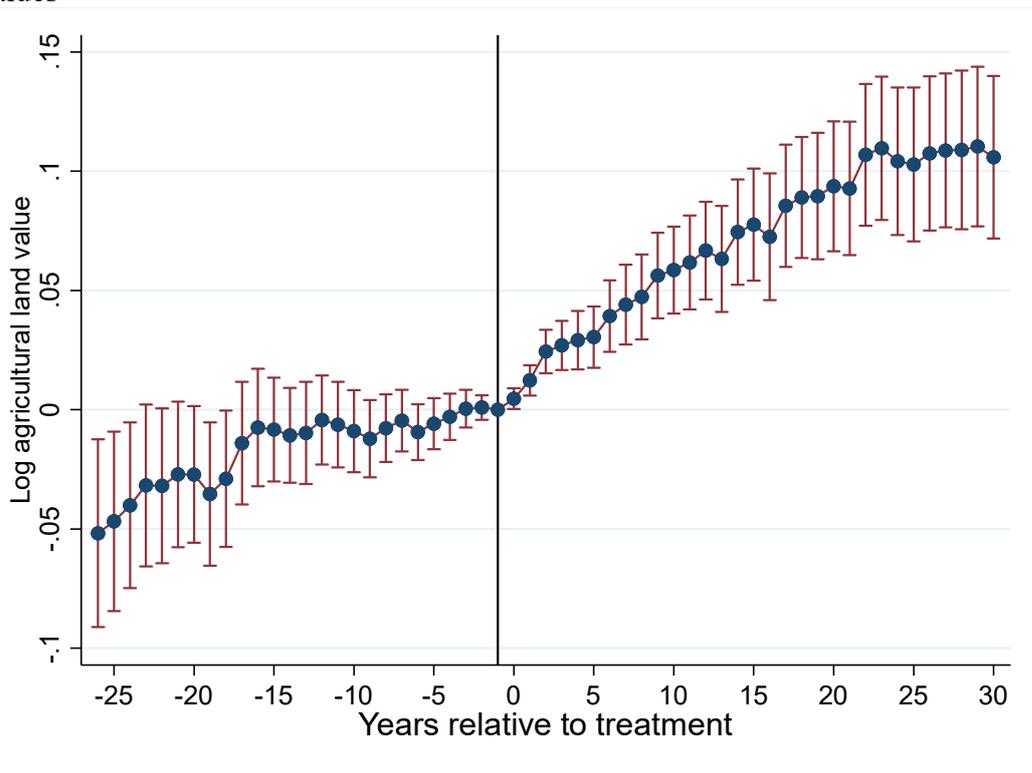

Figure 12A. Results from the de Chaisemartin and D'Haultfoeuille estimator: agricultural land values

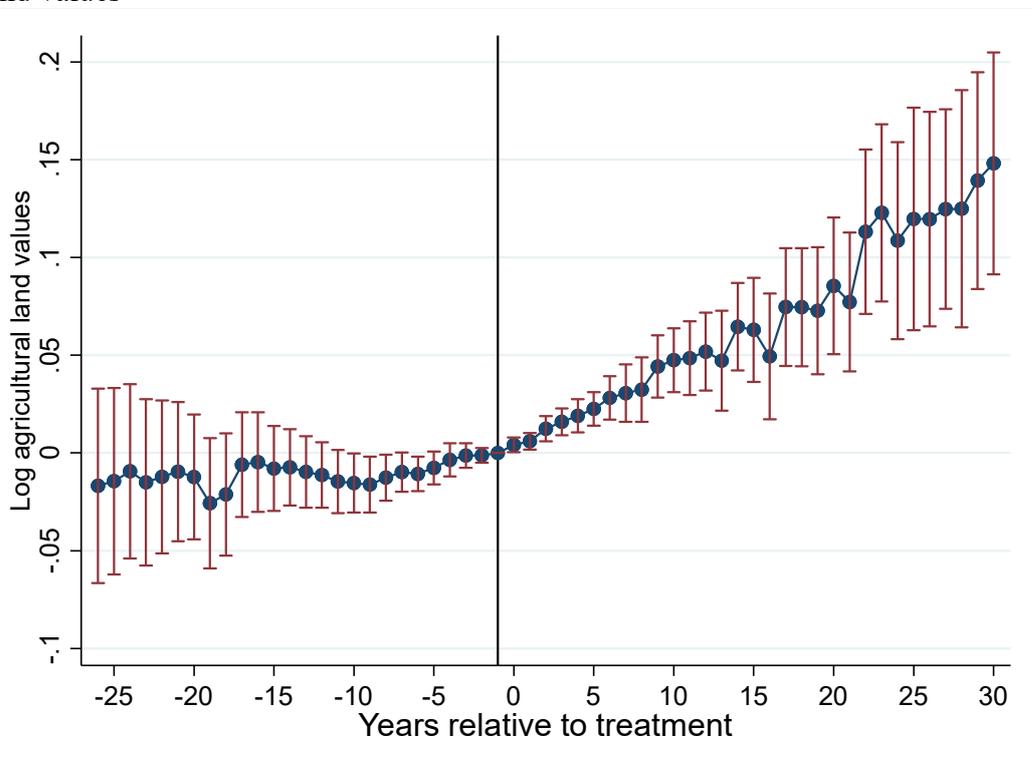

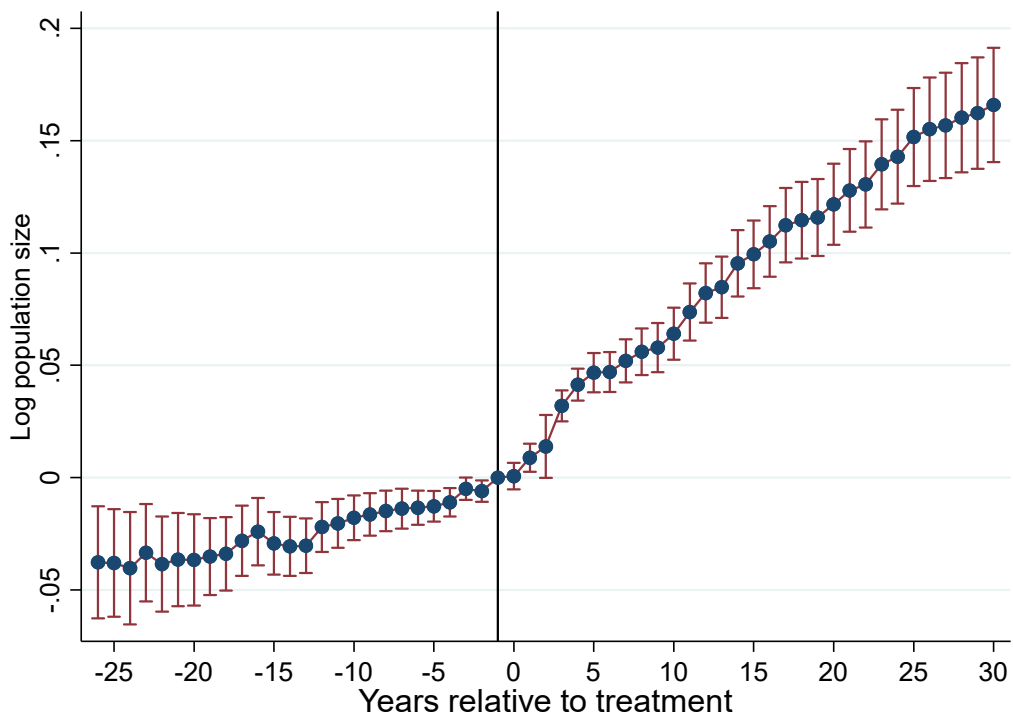

Figure 13A. Results from a TWFE specification with unrestricted dynamics: population size

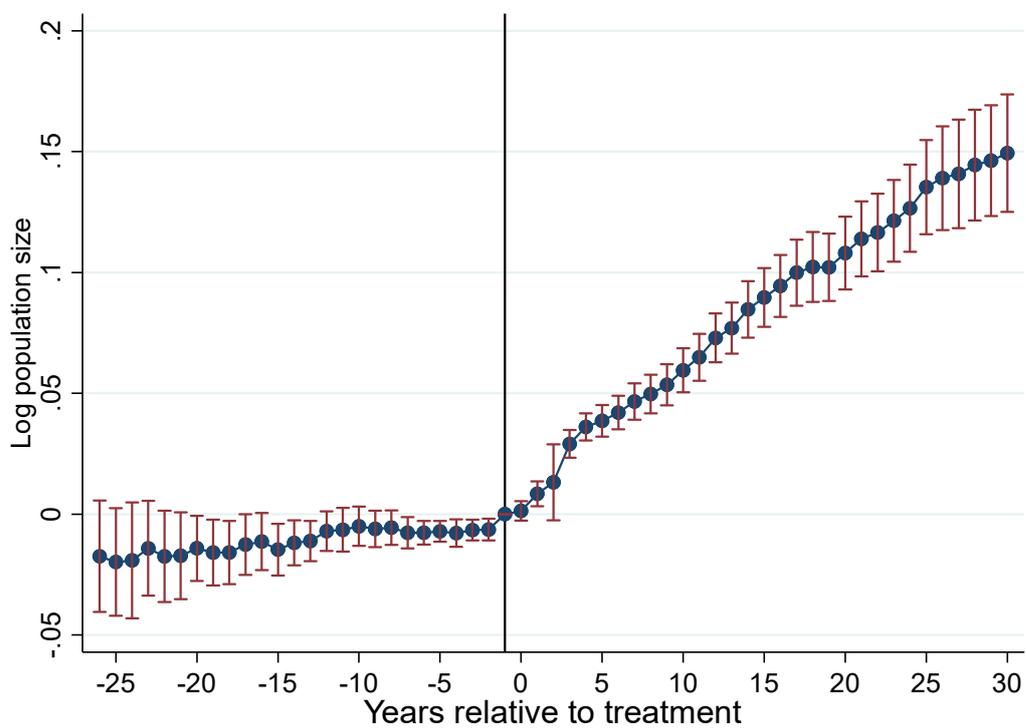

Figure 14A. Results from the de Chaisemartin and D'Haultfoeuille estimator: population size

# References for the appendix

Borusyak, Kirill, Xavier Jaravel and ann Spiess (2021). "Revisiting Event Study Designs: Robust and Efficient Estimation." mimeo

Cattaneo, Matias D., Richard Crump, Max Farrell and Yingjie Feng (1919). "On Binscatter" mimeo.

de Chaisemartin, Clement, and Xavier d'Haultfoeuille (2020). "Two-way fixed effects estimators with heterogeneous treatment effects." American Economic Review 110.9: 2964-96.

de Chaisemartin, Clement, and Xavier d'Haultfoeuille (2021). "Difference-in-Differences Estimators of Intertemporal Treatment Effects." Working paper

Clarke, Damian, and Kathya Tapia Schythe (2020), "Implementing the Panel Event Study", forthcoming in Stata Journal

Donaldson, David (2018). "Railroads of the Raj: Estimating the Impact of Transportation Infrastructure." American Economic Review 2018, 108(4-5): 899–934.

Donaldson, David and Richard Hornbeck (2016). Railroads and American Economic Growth: A "Market Access" Approach, The Quarterly Journal of Economics, Volume 131, Issue 2, pages 799–858.

Heblich, Stephan, Stephen J Redding, Daniel M Sturm (2020), "The Making of the Modern Metropolis: Evidence from London," *The Quarterly Journal of Economics*, Volume 135, Issue 4, pages 2059–2133,

Hornung, Erik, (2015). "Railroads and Growth in Prussia," Journal of the European Economic Association, Volume 13, Issue 4, 1, Pages 699–736

Schmidheiny, Kurt, and Sebastian Siegloch (2020). "On Event Studies and Distributed-Lags in Two-Way Fixed Effects Models: Identification, Equivalence, and Generalization". CEPR Discussion Paper 13477.

Sun, Liyang, and Sarah Abraham (2021). "Estimating dynamic treatment effects in event studies with heterogeneous treatment effects." Journal of Econometrics.